\documentclass[12pt,preprint]{aastex}

\bibpunct{(}{)}{;}{a}{}{,}

\slugcomment{May 20, 2008}
\shorttitle{New 3D$-$1D corrected abundances of HE~1327$-$232}
\shortauthors{A. Frebel et al.}

\begin{document}
\title{HE~1327$-$2326, an unevolved star with $\mbox{[Fe/H]}<-5.0$.\\
  II. New 3D$-$1D corrected abundances from a VLT/UVES spectrum\altaffilmark{1}}

\author{
Anna Frebel\altaffilmark{2},
Remo Collet\altaffilmark{3},
Kjell Eriksson\altaffilmark{4},
Norbert Christlieb\altaffilmark{4}
and Wako Aoki\altaffilmark{5, 6}}

\altaffiltext{1}{Based on observations collected at the European
Southern Observatory, Paranal, Chile (Proposal ID 075.D-0048).}

\altaffiltext{2}{McDonald Observatory, The University of Texas at
Austin, 1 University Station, C1400, Austin, TX 78712-0259;
anna@astro.as.utexas.edu}

\altaffiltext{3}{Max-Planck-Institut f{\"u}r Astrophysik, Karl-Schwarzschild-Str.~1,
Postfach 1317, D$-$85741 Garching b. M{\"u}nchen, Germany}

\altaffiltext{4}{Department of 
Astronomy and Space Physics, Uppsala University, Box 515, SE-751-20 Uppsala, 
Sweden; norbert@astro.uu.se} 

\altaffiltext{5}{National Astronomical Observatory of Japan, 2-1-21 Osawa,
Mitaka, Tokyo, 181-8588 Japan}

\altaffiltext{6}{Department of Astronomical Science, The Graduate
University of Advanced Studies, Mitaka, Tokyo, 181-8588 Japan}

\begin{abstract}
We present a new abundance analysis of HE~1327$-$2326, the currently
most iron-poor star, based on observational data obtained with
VLT/UVES. We correct the 1D LTE abundances for 3D effects to provide
an abundance pattern that supersedes previous works, and should be
used to observationally test current models of the chemical yields of
the first-generation SNe. Apart from confirming the 1D LTE abundances
found in previous studies before accounting for 3D effects, we make
use of a novel technique to apply the 3D$-$1D corrections for CNO
which are a function of excitation potential and line strength for the
molecular lines that comprise the observable CH, NH, and OH
features. We find that the fit to the NH band at 3360\,{\AA} is
greatly improved due to the application of the 3D$-$1D
corrections. This may indicate that 3D effects are actually observable
in this star. We also report the first detection of several weak Ni
lines. The cosmologically important element Li is still not detected;
the new Li upper limit is extremely low, $A({\rm Li})<0.62$, and in
stark contrast with results not only from WMAP but also from other
metal-poor stars. We also discuss how the new corrected abundance
pattern of HE~1327$-$2326 is being reproduced by individual and
integrated yields of SNe.

\end{abstract}


\section{Introduction}
The interpretation of the chemical abundance patterns of the most
metal-poor Galactic stars provides important clues to our
understanding of the early Universe and the chemical evolution of the
Milky Way. To advance our knowledge about this early time the observed
stellar signatures have to be compared with theoretical models of the
first cosmic chemical enrichment events. Currently, several groups
\citep{UmedaNomotoNature,chieffi_limongi2002, meynet2002, meynet2005,
heger2008} are working on different models of Population\,III
(Pop\,III) supernovae (SNe). The elemental yields of these SN
calculations are used, among other things, for a comparison with
individual old metal-poor stars as well as groups of metal-poor stars
with similar chemical patterns. These stars are believed to have
formed from material exclusively enriched during these early SN
events. It can thus be tested whether just a single SN was responsible
for the chemical signature in what could be some of the first low-mass
stars that formed in the Universe.

The recent discoveries of two stars with iron abundances of
$\mbox{[Fe/H]}\footnote{Where \mbox{[A/B]}$ = \log(N_{\rm A}/N_{\rm
B}) - \log(N_{\rm A}/N_{\rm B})_\odot$ for the number N of atoms of
element A and B and $\odot$ refers to the Sun}<-5$
\citep{HE0107_Nature, HE0107_ApJ, HE1327_Nature, Aokihe1327} have
already proved to be ideal test stars for this scenario because they
likely formed very early on in the Universe. \citet{UmedaNomotoNature}
first explained the abundance pattern of HE~0107$-$5240 with the
yields by one single SN undergoing a mixing and fallback mechanism. 
Subsequent work by \citet{iwamoto_science} explained HE~1327$-$2326's
signature in the same manner.  However, constructing such a SN model
requires as many observed abundances as possible-- it is a challenge
for this star, as it is extremely iron-deficient, and any spectral
lines are very weak (intrinsically, and because of the effective
temperature of 6180\,K) and difficult to detect.

In this paper, we thus present new abundance measurements of
HE~1327$-$2326, the currently most iron-poor star known (Frebel et
al. 2005; Aoki et al. 2006).  In 2005, a new VLT spectrum was
obtained. The initial discovery spectrum had been taken with the
Subaru telescope. The new data have higher $S/N$ and thus already
facilitated the determination of the O abundance of HE~1327$-$2326
from very weak near-UV OH lines (Frebel et al. 2006). The O
abundance provided a crucial ingredient for the interpretation of the
overall chemical abundance pattern (e.g., \citealt{iwamoto_science,
meynet2005}).

We now use these new VLT data to i) re-determine the abundances of all
elements obtained from the Subaru spectrum, ii) to search for
(additional) lines of new and already detected elements in the star,
iii) to obtain tighter upper limits of existing estimates, and iv) to
correct all abundances arising from our 1D model stellar atmosphere
analysis for 3D effects \citep{collet06}. For some elements, these
effects are quite significant. Our measurements, and in particular the
new detection of Ni, will provide observational constraints on the
yields of the first generations of SNe. Hence, the enrichment scenario
of HE~1327$-$2326 can be further explored, as well as the origins of
other extremely metal-poor stars.


\section{Observations}
A description of the spectroscopic observations can be found in Frebel et
al. (2006). We remind the reader that due to the choice of settings
the wavelength coverage of the new data is limited to three regions:
3050--3870\,{\AA}, 4780--6805\,{\AA}, and 5720--9470\,{\AA}. The data
were reduced with the REDUCE package \citep{reduce} and subsequently
rebinned. The signal-to-noise ($S/N$) ratios of different wavelength
regions are listed in Table~\ref{Tab:SN}.

\section{Stellar Parameters and Model Atmospheres}
We adopt an effective temperature of $T_{\rm eff}=6180$\,K, as
determined in our previous analyses \citep{HE1327_Nature,
Aokihe1327}. Back then it was not possible, however, to distinguish whether
HE~1327$-$2326 would be a subgiant or a main-sequence star. In the
meantime, \citet{korn_fs3} carried out a new analysis of the Ca\,I/II
ionization equilibrium, which functions as a surface-gravity
indicator. As Ca\,I line formation is prone to departures from LTE
\citep{Mashonkina_ca}, detailed calculations have been necessary to
determine the magnitude of non-LTE effects.  Based on the results of a
combined analysis of Ca\,I/II ionization equilibrium and Balmer lines,
\citet{korn_fs3} conclude that HE~1327$-$2326 is more likely a
subgiant. Hence, for the present work, we will consider the star to
be in this evolutionary stage ($\log g= 3.7$).

For the 1D abundance analysis, we use a customary 1D LTE {\sc marcs}
model stellar atmosphere (B. Gustafsson et al. 2008, in prep.).  The
model has been constructed with $T_{\rm eff}=6180$\,K, $\log g= 3.7$,
and $v_{micr}=2.0$\,km/s.  The complete abundance pattern of the star
as derived by \citet{Aokihe1327} has been taken into account in the
actual modeling.  Incidentally, \citet{Aokihe1327} also showed that
an analysis based on a Kuruzc model with a solar chemical composition
scaled down to $\mbox{[Fe/H]}=-5$ returns a very similar abundance
pattern for HE~1327$-$2326 (within 0.1-0.2\,dex). The mixing-length
theory (MLT) formulation for the {\sc marcs} model adopted here comes
from \citet{henyey06}.

Recently, \citet{collet06} presented 3D$-$1D corrections to the
elemental abundances of HE~1327$-$2326 derived from individual atomic
as well as molecular lines.  Such corrections were computed under the
assumption of local thermodynamic equilibrium (LTE) by means of a
differential 3D$-$1D analysis based on a 3D hydrodynamical and a 1D
{\sc marcs} model atmosphere with the same stellar parameters; the
authors then applied the results of their 3D$-$1D analysis to the
abundances derived by \citet{Aokihe1327} and \citet{o_he1327} with a
1D analysis to obtain the 3D$-$1D corrected chemical composition of
{HE~1327$-$2326}.  In the present work, we carry out a differential
3D$-$1D LTE abundance analysis for the newly observed atomic lines
following the same procedure as in \citet{collet06}.  With the
availability of such corrections, in this paper we focus on correcting
the abundances of HE~1327$-$2326 for 3D effects to provide the most
accurate as possible stellar abundances.

\section{Differential 3D$-$1D abundance analysis}
We use convection simulation of a metal-poor turnoff star (T$_{\rm
eff}{\simeq}6200$~K, $\log g=4.04$~[cgs], and solar chemical
composition scaled down to $\rm{[Fe/H]}=-3$) previously generated by
\citet{asplund01} as a time-dependent 3D hydrodynamical model
atmosphere to study the formation of the atomic and molecular features
observed in HE~1327$-$2326 under the assumption of LTE.  The surface
convection simulation was carried out with the 3D, time-dependent,
compressible, explicit, Eulerian, radiative-hydrodynamical code by
\citet{nordlund82,stein98}.  The physical domain of the simulation is
a cubic portion of stellar surface about $21{\times}21{\times}7$~Mm in
size, large enough to cover about ten granules horizontally and twelve
pressure scales in depth.  The domain is discretized on a mesh with
$100{\times}100{\times}82$ numerical resolution.  In terms of
continuum optical depth at $\lambda=5000$~{\AA}, the simulation
extends from $\log{\tau_{\rm{5000}}}\la-5$ down to
$\log{\tau_{\rm{5000}}}\ga6$.  Open boundaries are employed vertically
and periodic ones horizontally.

\begin{figure}
 \begin{center}
  \includegraphics[height=9cm, clip=true,bbllx=0, bblly=0, bburx=540, bbury=360]{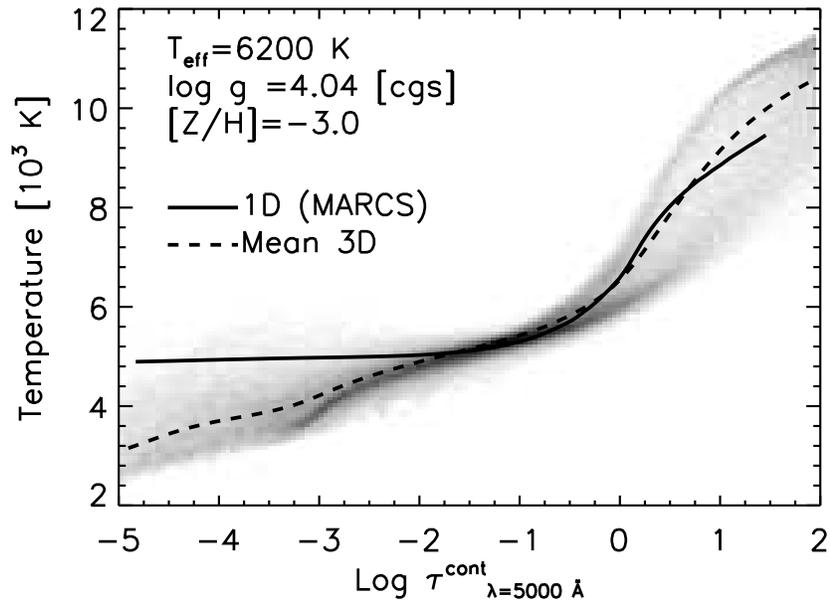} 
   \figcaption{ \label{optical_depth}
   Temperature structure the upper layers of the 3D hydrodynamical
   simulation adopted in the present 3D$-$1D differential 
   abundance analysis.
   \textit{Gray shaded area}: temperature distribution as a function of optical 
   depth at $5000$~{\AA} in the 3D model. 
   Darker areas indicate temperature values with higher probability. 
   \textit{Dashed line}: mean temperature stratification of the 
   simulation (averaged over surfaces of constant optical depth at 
   $5000$~{\AA}). \textit{Solid line}: temperature stratification 
   of the 1D {\sc marcs} model constructed for the same stellar parameters.}
 \end{center}
\end{figure}

The adopted gravity of $\log g=4.04$ in the 3D simulation is
intermediate between what until recently were considered the two best
estimates of HE~1327$-$2326's surface gravity, i.e., $\log g=3.7$
(subgiant solution) or $\log g=4.5$ (dwarf).  In solar-like stars,
absolute elemental abundances derived from lines of minority species
are essentially insensitive to the exact value of surface gravity
\citep[see also Tab.~9 and~10 in][]{Aokihe1327}.  On the contrary,
abundances inferred from lines of majority species (e.g., Fe~II) and
molecules are more surface gravity dependent.  In practice, such
surface gravity sensitivity of the derived abundances also changes
depending on whether one considers a 3D or 1D model atmosphere for the
spectral line formation calculations.  Therefore, rather than directly
determining the chemical composition of HE~1327$-$2326 with the
exclusive use of the above 3D model, we instead compute
\emph{differential} abundances with respect to the results of a 1D
abundance analysis based on a 1D {\sc marcs} model atmosphere
\citep{gustafsson75,asplund97} with identical stellar parameters,
input data, and chemical composition as the convection simulation.
Figure~\ref{optical_depth} shows the temperature structure of the 3D
model atmosphere adopted here as a function of optical depth; the
temperature stratification of the corresponding 1D {\sc marcs} model
atmosphere is also plotted for comparison.  For relatively small
changes in terms of surface gravity, such \emph{differential} 3D$-$1D
should in fact be only very marginally sensitive to $\log g$.

While both the 3D simulation and the 1D {\sc marcs} model atmosphere
are constructed for a solar chemical composition scaled down to
$\rm{[Fe/H]}=-3$, in the line formation calculations we assume the
chemical abundance pattern to be the same as for HE~1327$-$2326
\citep{Aokihe1327} when computing ionization and molecular equilibria
and continuous opacities; as a rule, only the abundances of the trace
elements are varied when calculating line opacities.  In fact, using a
scaled solar chemical composition with $\rm{[Fe/H]}=-3$ would lead to
overestimate the abundance of elements with low ionization potentials
and, in turn, the electron density, therefore affecting ionization
balance, continuous opacities, and, ultimately, line strengths.  In
addition, when computing background opacities and electron density, we
adopt C, N, and O abundances midway between the values derived by
\citet{Aokihe1327} and \citet{o_he1327} for the subgiant and dwarf
solutions: this is done in order to compensate for the particular
choice of surface gravity in the 3D simulation and corresponding 1D
model.

We compute flux profiles for all the spectral lines from neutral and
singly-ionized metals presently observed in HE~1327$-$2326.  We also
consider a set of ``fictitious'' CH, NH, and OH (see
e.g., \citealt{collet06,collet07}) lines with varying lower excitation
potentials ($0$ to $3.5$~eV) and $\log{gf}$ values within the range
typical for the observed molecular bands in this study.  From the full
3D simulation we select a 60~minutes long (stellar time) sequence of
30 snapshots equidistant in time.  We decrease the horizontal
resolution of the simulation from $100{\times}100$ to $50{\times}50$
to reduce the computational load and time of the 3D line formation
calculations; we also interpolate all simulation snapshots to a finer
depth-scale in order to increase the resolution of the atmospheric
layers ($\log{\tau_{\rm{5000}}}{\la}2.5$) to improve the numerical
accuracy of the radiative transfer solution.  For each line we solve
the radiative transfer equation for all grid-points across the
horizontal plane and along 33 directions (4 $\mu$-angles, 8
$\phi$-angles, and the vertical), after which we perform a
disk-integration and a time-average over all selected 3D snapshots.
We use 70 wavelength points to resolve each line profile.  The source
function for lines and continuum is approximated with the Planck
function at the local temperature and scattering is treated as true
absorption.  Continuous opacities come from the Uppsala package
\citep[][and subsequent updates]{gustafsson75}, the equation-of-state
from \citet[][and subsequent updates]{mihalas88}, the partition data
for atoms and ions from \citet{irwin81} and for molecules from
\citet{sauval84}.  The same radiative transfer solver and input data
are also used for spectral line formation calculations with the 1D
model atmosphere. In addition, for the 1D case, we adopt a
micro-turbulence of $\xi=1.6$\,km\,s$^{-1}$ for the turnoff star.  We
stress that no micro- nor macro-turbulence parameters enter the 3D
spectral line synthesis calculations: only the velocity fields
predicted by the actual 3D simulation are used to reproduce
non-thermal line broadening and asymmetries associated with convective
Doppler shifts.

For each line, we quantify the impact of the 3D model in the analysis, 
by varying the abundance of the trace element
independently in the 1D and 3D line formation calculations
until the measured equivalent width is matched:
the difference between the 3D and 1D derived abundances then
defines the 3D$-$1D abundance correction for the given line.

Before proceeding, we caution that the actual 3D$-$1D abundance
corrections depend in part on the choice of 1D model atmosphere in the
differential abundance analysis.  Ideally, one should compute the
differential corrections using a 1D model atmosphere that relies on
exactly the same input physics and treatment of radiative transfer as
the convection simulation.  Although the input physics of the {\sc
marcs} model atmosphere adopted here does not depart substantially
from the one used in the 3D case, some differences in practice still
exist.  In particular, contrary to the 3D hydrodynamical simulation,
scattering is treated correctly as such and not as true absorption in
the 1D {\sc marcs} model.  Also, the two kinds of models rely on
slightly different equations-of-state. Finally, while the sources of
continuous and line opacities are namely the same
\citep{gustafsson75,kurucz92,kurucz93}, during the 3D simulation
opacities are grouped in four opacity bins to ease the computational
burden of the radiative transfer calculations.  To estimate what the
effect of the particular choice of model atmosphere is on the 3D$-$1D
corrections, we also performed a 3D$-$1D differential abundance
analysis based on a 1D {\sc atlas} model atmosphere \citep{kurucz93}
with same stellar parameters and chemical composition as the 3D model.
Because of the slightly higher temperature of the 1D {\sc atlas} model
atmosphere in line formation regions, the resulting 3D$-$1D abundance
corrections for all elements are systematically shifted $0.05$ to
$0.1$\,dex downwards (i.e., toward more negative values) with respect
to the differential analysis based on the 1D {\sc marcs} model. We
finally note that there is no dependence of the 3D$-$1D corrections on
the adopted micro-turbulence of the 1D model (see
\citealt{Aokihe1327}), nor any significant dependency on
MLT. Molecular lines, for which the 3D$-$1D corrections are the largest,
form in the upper layers of the photosphere where the type of
mixing-length prescription is not expected to have an effect on the
atmospheric structure.

\section{Chemical Abundances}
\begin{figure}
 \begin{center}
  \includegraphics[height=9cm, clip=true,bbllx=60, bblly=417, bburx=480, bbury=711]{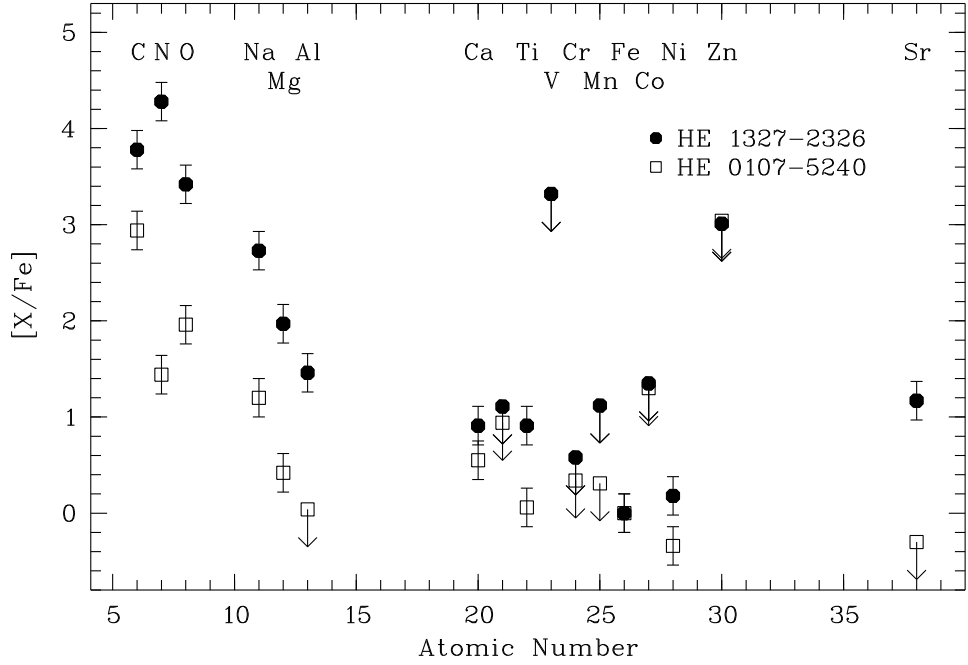}
  \figcaption{\label{abund_pattern} Abundance pattern of
  HE~1327$-$2326 (\textit{filled circles}) in comparison with that of
  HE~0107$-$5240 (\textit{open squares}). Typical $1\,\sigma$ errors of
  $0.2$\,dex are shown. Upper limits are indicated by an arrow. The 3D
  abundances for HE~0107$-$5240 are taken from \citet{collet06}. }
\end{center}
\end{figure}

We now describe the abundances of individual elements as far as the
new analysis is concerned. Details on the 3D$-$1D corrections and our
new technique to apply those to abundances derived from molecular
features are given where appropriate. The new abundance pattern is
summarized in Table~\ref{Tab:Abundances}, and in
Figure~\ref{abund_pattern} our results for HE~1327$-$2326 are compared
with those for HE~0107$-$5240, the other star with
$\mbox{[Fe/H]}<-5$. We also refer the reader to \citet{Aokihe1327};
they presented a detailed discussion of the HE~1327$-$2326's chemical
pattern as well as a comparison with other metal-poor stars. A
detailed uncertainty analysis for the abundances of the star was also
presented (their Table~10) and will not be repeated here.

\subsection{Line Data and Analysis Techniques}
Atomic line data for the abundance analysis have been taken from
\citet{Aokihe1327} and references therein, substituted with lines from
\citet{HE0107_ApJ} and references therein, and from the Vienna Atomic Line
Database (VALD; \citealt{vald}). Molecular line data have been taken
from B. Plez (priv. communication) for CH and
\citet{gillis_ohlinelist} for OH. For NH we are using the linelist of
\citet{kurucz_nh}, but add a 0.4\,dex to the gf-values to fit the
solar spectrum (see \citealt{Aokihe1327} for details). For CH and OH we use
dissociation energies of 3.47\,eV and 4.39\,eV, respectively
\citep{CH_OH_diss_en}. For NH we employ 3.37\,eV \citep{NH_diss_en}.

Equivalent widths are measured from the spectrum by fitting Gaussian
profiles to the generally very weak lines.  Upper limits ($3\sigma$)
for equivalent widths are derived from $\sigma=w n_{pix}^{1/2}/(S/N)$
(where $w$ is the pixel width, $n_{pix}$ is the number of pixels
across the line (e.g., where a weak line would reach the continuum if
a line was present), and $S/N$ is per pixel; \citealt{bohlin}). The
measurements and upper limits for HE~1327$-$2326 are presented in
Table~\ref{Tab:Eqw}. Apart from hydrogen and molecular lines of CH,
NH, and OH we now find 44 atomic lines in the spectrum of eight
elements (Na, Mg, Al, Ca, Ti, Fe, Ni, Sr) and are able to determine
upper limits of nine additional elements. The previous analysis by
\citet{Aokihe1327} identified 22 atomic lines.

\subsection{Iron}

\begin{figure}
 \begin{center}
\includegraphics[height=15cm, clip=true,bbllx=30, bblly=248, bburx=540, bbury=744]{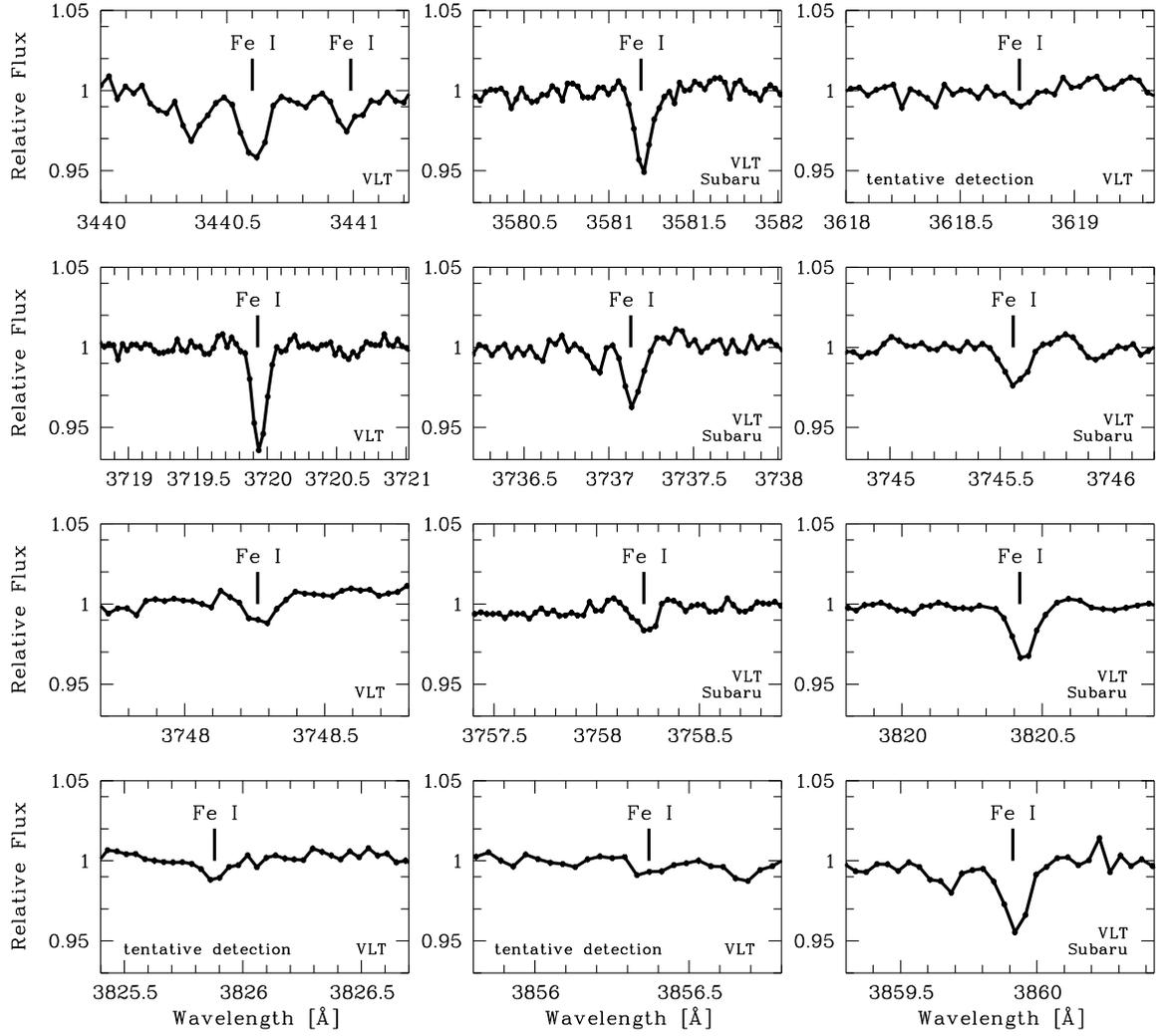} \figcaption{\label{feI_plot} All
  detected and tentatively detected Fe\,I lines in the VLT/UVES
  spectrum of HE~1327$-$2326. Equivalent width measurements can be
  found in Table~\ref{Tab:Eqw}.}
\end{center}
\end{figure}

In the Subaru spectrum seven weak Fe\,I lines were detected
\citep{Aokihe1327}. The strongest line was measured to have an
equivalent width of 6.8\,m{\AA}. As already reported in Frebel et
al. (2006) the new data confirm the presence of six of the seven lines
(the wavelength of one line is not covered by our VLT
spectrum). Additionally, we detect four new lines and report tentative
detections of another three.  This brings the total of detected lines
to 11 Fe\,I lines from which the metallicity is deduced. We note that
the average that includes the abundances of the tentative Fe\,I lines
agrees within 0.03\,dex with the average from the 11 detected
lines only. Figure~\ref{feI_plot} shows the detected and tentatively
detected 13 Fe\,I lines in the VLT spectrum.

Based on the new equivalent width measurements together with the one
from Aoki et al. (2006) for the line that is not covered, we find 
HE~1327$-$2326 to have a 1D LTE Fe abundance\footnote{This value has
essentially already been published in \citet{o_he1327}. We simply give
more details on the derivation of this value, and also report the
detection of additional Fe\,I lines.} of $\mbox{[Fe/H]} =-5.66$, the
same as the previous result reported in \citet{Aokihe1327}.

\begin{figure}
 \begin{center}
  \includegraphics[height=9cm, clip=true,bbllx=0, bblly=0, bburx=504, bbury=360]{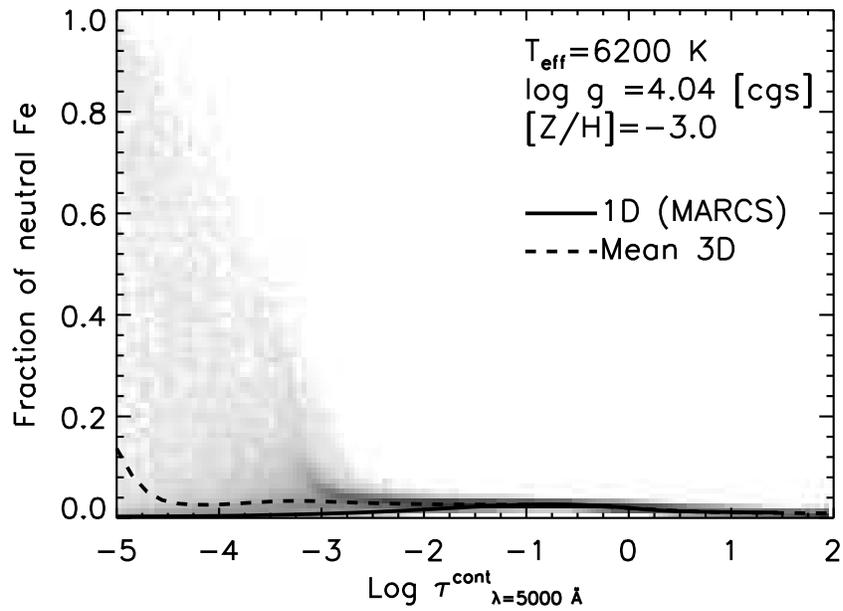}
  \figcaption{  \label{feifraction}
   Ratio of neutral to total iron number densities 
   ($n_{\rm{Fe\,I}}/n_{\rm{Fe}}$) as a function of
   optical depth at $5000$\,{\AA}. 
   \textit{Gray shaded area}: distribution of $n_{\rm{Fe\,I}}/n_{\rm{Fe}}$ 
   values in the 3D model atmosphere adopted here; 
   darker areas indicate values with higher probability. 
   Over-plotted are the curves for the mean 3D stratification 
   (\textit{dashed line}) and for the corresponding 1D {\sc marcs}
   model atmosphere (\textit{solid line}). }
 \end{center}
\end{figure}

\begin{figure}
 \begin{center}
  \includegraphics[height=9cm, clip=true,bbllx=0,bblly=0,bburx=504,bbury=360]{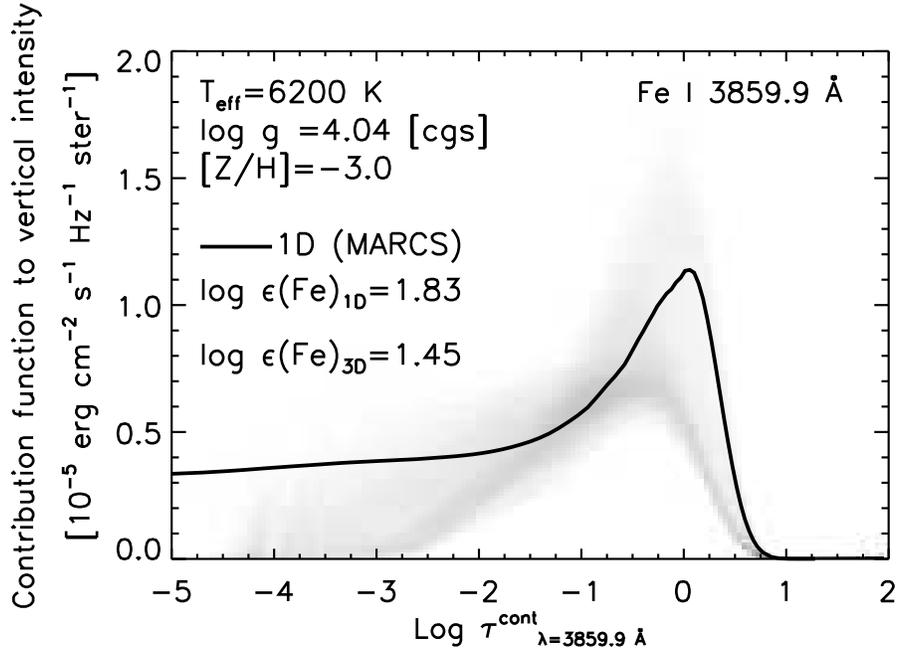}
   \figcaption{ \label{feicontrib}
    Contribution function to outgoing intensity in the vertical direction
    at line centre for the Fe\,I 3859.9\,{\AA} line
    as a function of optical depth.
    \textit{Gray shaded area}: distribution of the contribution function values
    in the 3D model atmosphere adopted here;
    darker areas indicate values with higher probability.
    Over-plotted (\textit{solid line}) is the contribution function
    computed with the corresponding 1D {\sc marcs} model. 
    The reported 3D and 1D Fe abundances are the ones reproducing
    the measured equivalent width of the line.}
\end{center}
\end{figure}

In this work we are now applying 3D$-$1D LTE corrections to our 1D LTE
abundances.  Within the LTE framework, the average 3D$-$1D Fe
abundance correction for Fe\,I lines is negative ($-0.24$;
\citealt{collet06}).  Thus, the Fe\,I abundance of HE~1327$-$2326 is
reduced to $\mbox{[Fe/H]}=-5.90$.  To illustrate why the 3D$-$1D Fe
abundance correction from Fe\,I lines is negative, in
Figure~\ref{feifraction} we compare the fraction of neutral to total
iron in the 3D and 1D model atmospheres.  While in the 1D model iron
is essentially all ionized to Fe\,II, in 3D a significant fraction of
iron is in neutral form in the upper photospheric layers because of
the cooler surface temperature stratification predicted by the 3D
metal-poor convection simulation (see e.g., \citealt{asplund01}).
Hence, at a given Fe abundance and under the assumption of LTE, Fe\,I
lines appear stronger in the 3D line formation calculations than they
do in 1D.  This also implies that the same equivalent width of a Fe\,I
line is matched by a lower Fe abundance in 3D than in 1D.
Alternatively, one can also show that the same Fe\,I line forms at
different depths in the 3D and 1D models.  Figure~\ref{feicontrib}
shows the contribution function
$\cal{C}_{\rm{I}}(\tau_\lambda)=S_{\lambda}(\tau_\lambda)\exp{(-\tau_\lambda)}$
to the outgoing intensity at line center in the vertical direction for
the Fe\,I line at $\lambda=3859.9$~{\AA} in the two models.  It is
apparent from the figure that, in 3D, the main contribution to the
outgoing vertical intensity for the Fe\,I line is given by layers
lying higher up in the atmosphere where the temperature differences
between 3D and 1D structures are larger.

We note that any possible non-LTE effects present within the 3D Fe\,I
line formation calculations are currently unexplored, but likely to be
positive \citep{asplund_araa}. For completeness, we remind the reader
that in previous works the measured Fe\,I abundance was corrected by
+0.2\,dex to account for 1D non-LTE effects in Fe\,I
(\citealt{collet06} predict for Fe\,I a much stronger departure from
LTE).  In principle, Fe\,II lines should be the preferred Fe abundance
indicator as they are usually expected to be less sensitive to
departures from LTE.  However, even the strongest Fe\,II lines are
expected to be extremely weak in HE~1327$-$2326 given its stellar
parameters and composition. In fact, no Fe\,II line could be detected
in either the Subaru or in the new VLT spectrum.  The 3D$-$1D corrected
upper limit is $\mbox{[Fe\,II/H]}<-5.40$. We can also place an upper
limit on the Fe\,I 3D non-LTE effect of $\lesssim0.5$. This is derived
under the assumption that there is an ionization equilibrium of Fe\,I
and Fe\,II if the surface gravity of $\log g=3.7$, as derived from an
isochrone, is correct, and that Fe\,II lines are not affected by
non-LTE. In the absence of Fe\,II lines, Ca is the only element
detectable in two ionization stages.  As mentioned above, this offers
the possibility of using Ca\,I/II as a surface-gravity indicator
\citep{korn_fs3}.

\subsection{CNO elements}

Carbon, nitrogen and oxygen abundances for HE~1327$-$2326 were derived
by \citet{Aokihe1327} (C and N; from the Subaru spectrum) and then by
\citet{o_he1327} (C and O; from the VLT spectrum). For the latter 1D LTE
analysis, corrections were suggested for all three elements to account
for 3D model atmosphere effects. The estimated 3D$-$1D corrections were
mostly based on works described in \citet{asplund_araa} and references
therein.

\begin{figure*}
 \begin{center}
 \resizebox{\hsize}{!}{
  \includegraphics{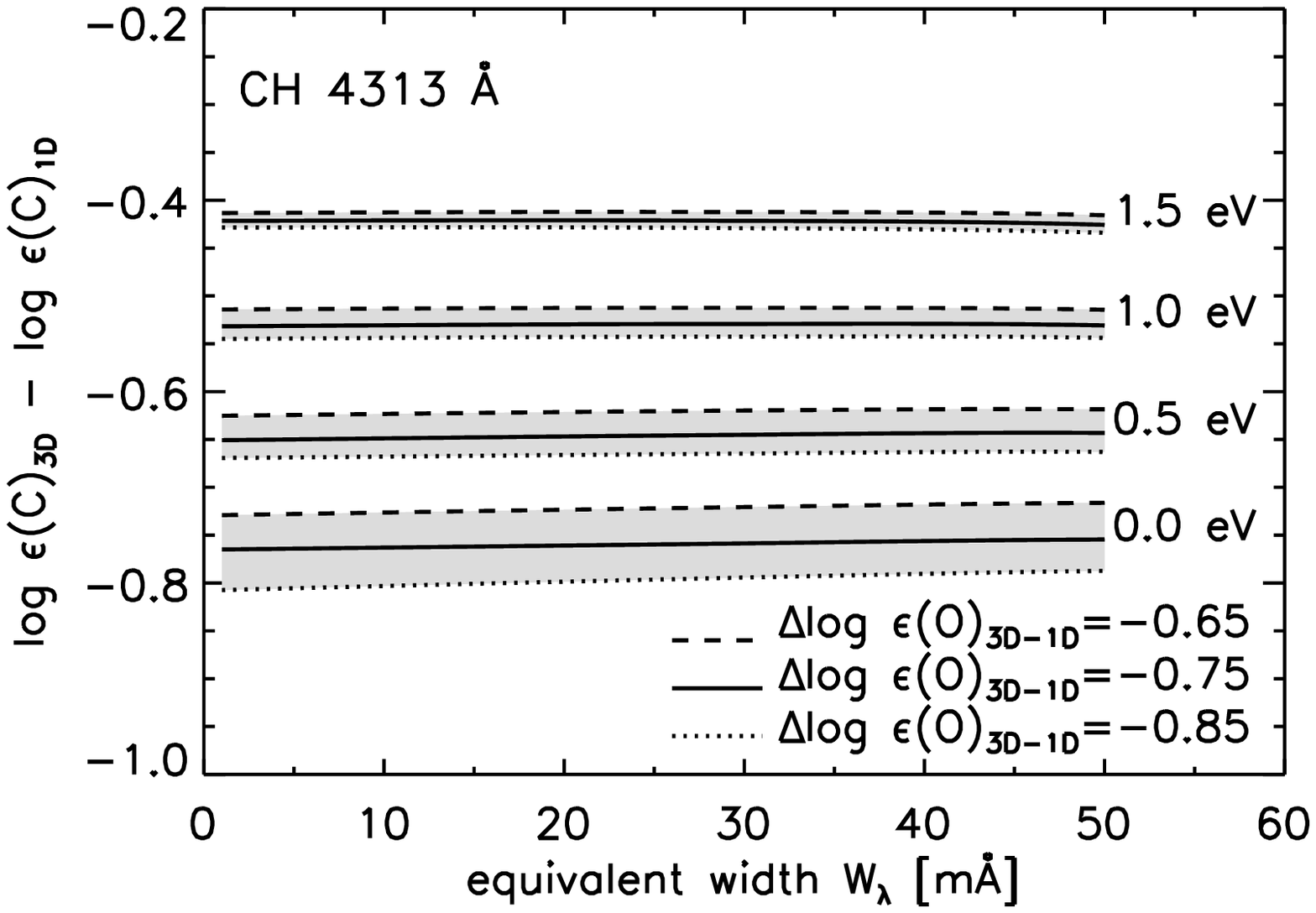}
  \includegraphics{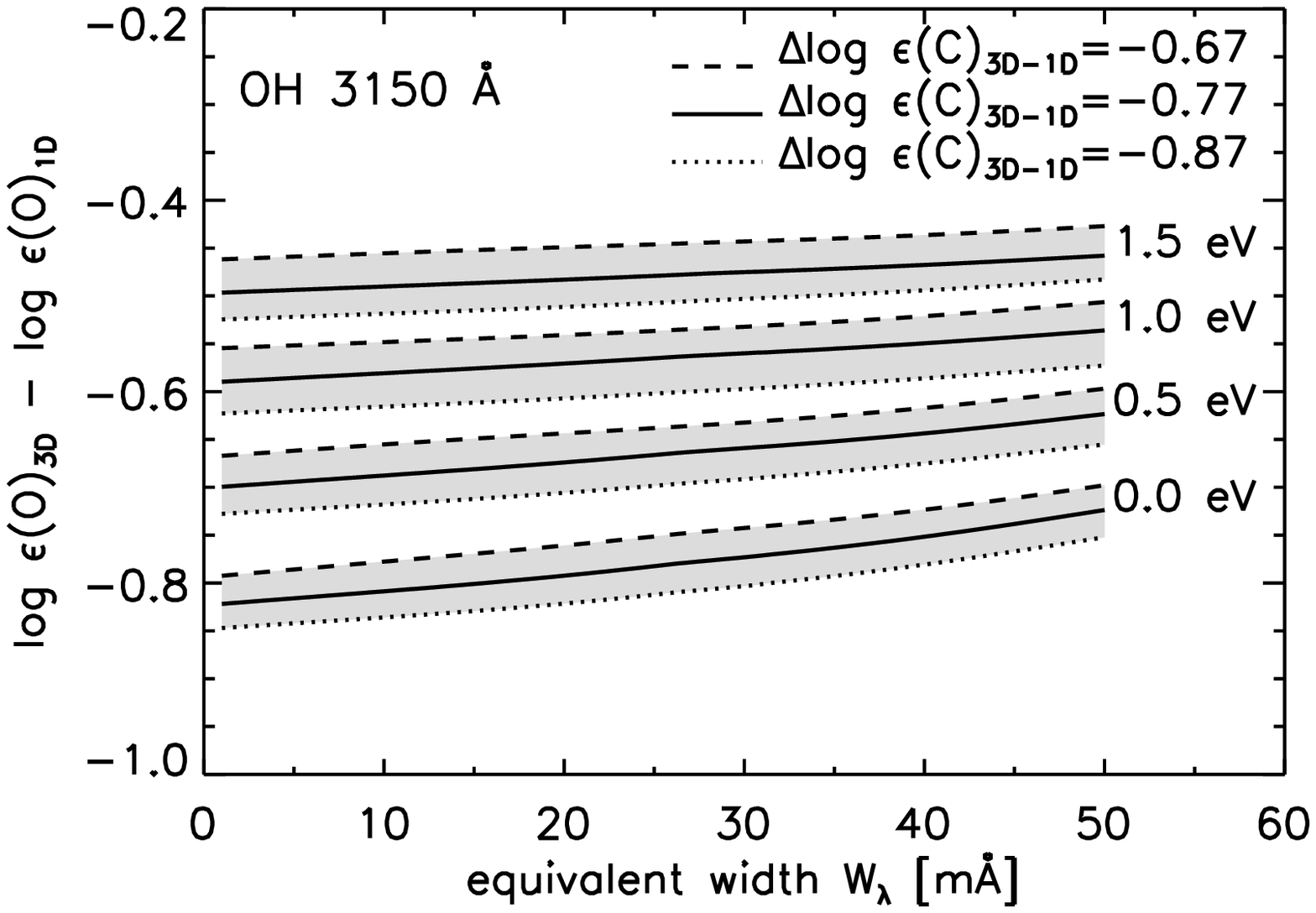} }
  \figcaption{  \label{ch_oh_corr}
  3D$-$1D LTE corrections to the C (\textit{left panel}) and O abundances 
  (\textit{right panel}) as derived from fictitious CH and OH lines. 
  Carbon and oxygen abundance corrections are plotted 
  as a function of line strength for selected values of the 
  lower excitation potential ($0$ to $1.5$\,eV) and for different values of
  3D O and C abundances respectively. }
\end{center}
\end{figure*}

\begin{figure}
 \begin{center}
  \includegraphics{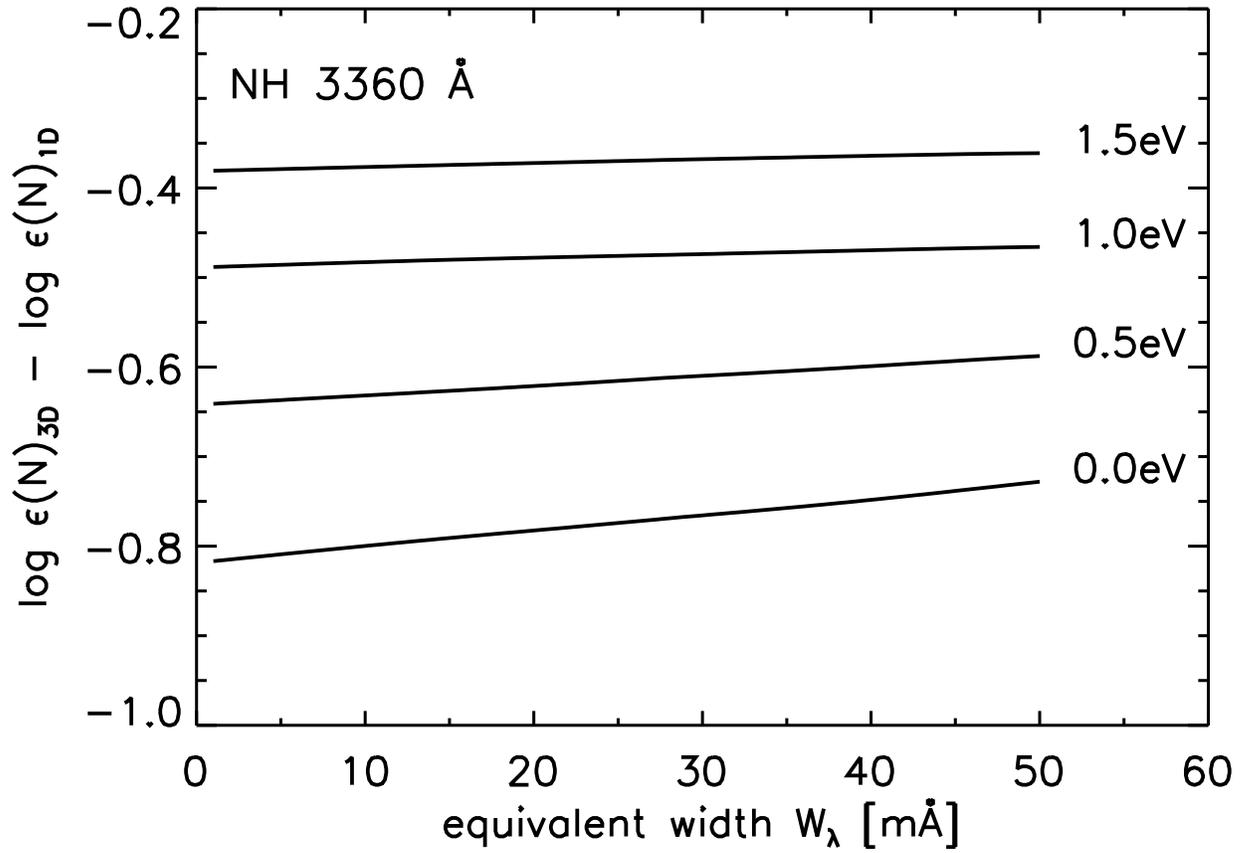}
  \figcaption{  \label{nh_corr}
  3D$-$1D LTE corrections to the N abundance as derived from
  fictitious NH lines. The corrections are plotted 
  as a function of line strength for different values of the 
  lower excitation potential.}
\end{center}
\end{figure}

Molecular features consist of many different lines with a variety of
excitation potentials.  As 3D$-$1D abundance corrections depend in
general on the excitation potentials of the contributing lines, the
final 3D$-$1D correction to the 1D abundance of the trace elements in the
molecular features has to be an appropriate average of the 3D$-$1D
abundance corrections derived for individual lines.  \citet{collet06}
computed 3D$-$1D corrections for a few fictitious lines of CH, NH, and
OH with selected excitation potentials.  By using fictitious lines,
the 3D$-$1D corrections to the CNO abundances could be studied without
resorting to a full spectrum synthesis with a 3D model atmosphere.
Here we extend the study by \citet{collet06} to consider a more
comprehensive range of lower excitation potentials and $\log{gf}$
values of the CH, NH, and OH lines.  Figures~\ref{ch_oh_corr}
and~\ref{nh_corr} show the 3D$-$1D LTE corrections to the CNO
abundances derived from the molecular lines as a function of line
strength and excitation potential.  In the case of CH and OH lines,
3D$-$1D LTE C and O abundances are also plotted as a function of the
corrections to oxygen and carbon abundances, respectively.  The
magnitudes of abundance corrections derived from molecular lines
depend in fact not only on the different photospheric structures of
the 3D and 1D model atmospheres but also on the overall chemical
composition and details of the molecular equilibrium.  In particular,
the line strengths of CH and OH lines are sensitive to both the carbon
\emph{and} oxygen abundances assumed in the spectral line formation
calculations because of the competing formation of CO molecules.
Temperatures in the upper photospheric layers of the metal-poor 3D
model atmosphere adopted here are on average considerably lower than
in the corresponding 1D model.  At low temperatures (${\la}3700$~K),
significant fractions of C and O become locked in CO molecules.  Thus,
for instance, if one were to increase slightly the \emph{oxygen}
abundance in the line formation calculations with the metal-poor 3D
model, that would then lead to the formation of more CO and therefore
reduce the amount of \emph{carbon} available for other molecules such
as CH. Hence, at a fixed carbon abundance, but with higher oxygen
abundance, CH lines would become \emph{weaker}, and the associated
3D$-$1D LTE C abundance corrections would in turn be \emph{larger}
(i.e., more negative).  The same argument holds for the variation of
the equivalent widths of OH lines as a function carbon abundance.

In order to apply the 3D$-$1D corrections to the CNO abundances in a
more suitable way, we developed a new technique. We first interpolate
in between the 3D$-$1D abundance corrections given as a function of
excitation potential and line-strength to obtain appropriate
corrections for all individual lines that comprise the molecular
feature of interest.  The abundance corrections so computed (in
\emph{dex}) are then \emph{subtracted} (the negative corrections) from
the $\log{gf}$ values for all lines in our linelist.  The observed
spectrum is then re-synthesized with the new ``3D-adjusted'' linelist.
We call this procedure ``3D-aided spectrum synthesis''.

\begin{figure}
 \begin{center}
\includegraphics[height=16cm, clip=true,bbllx=50, bblly=164,
  bburx=526, bbury=714]{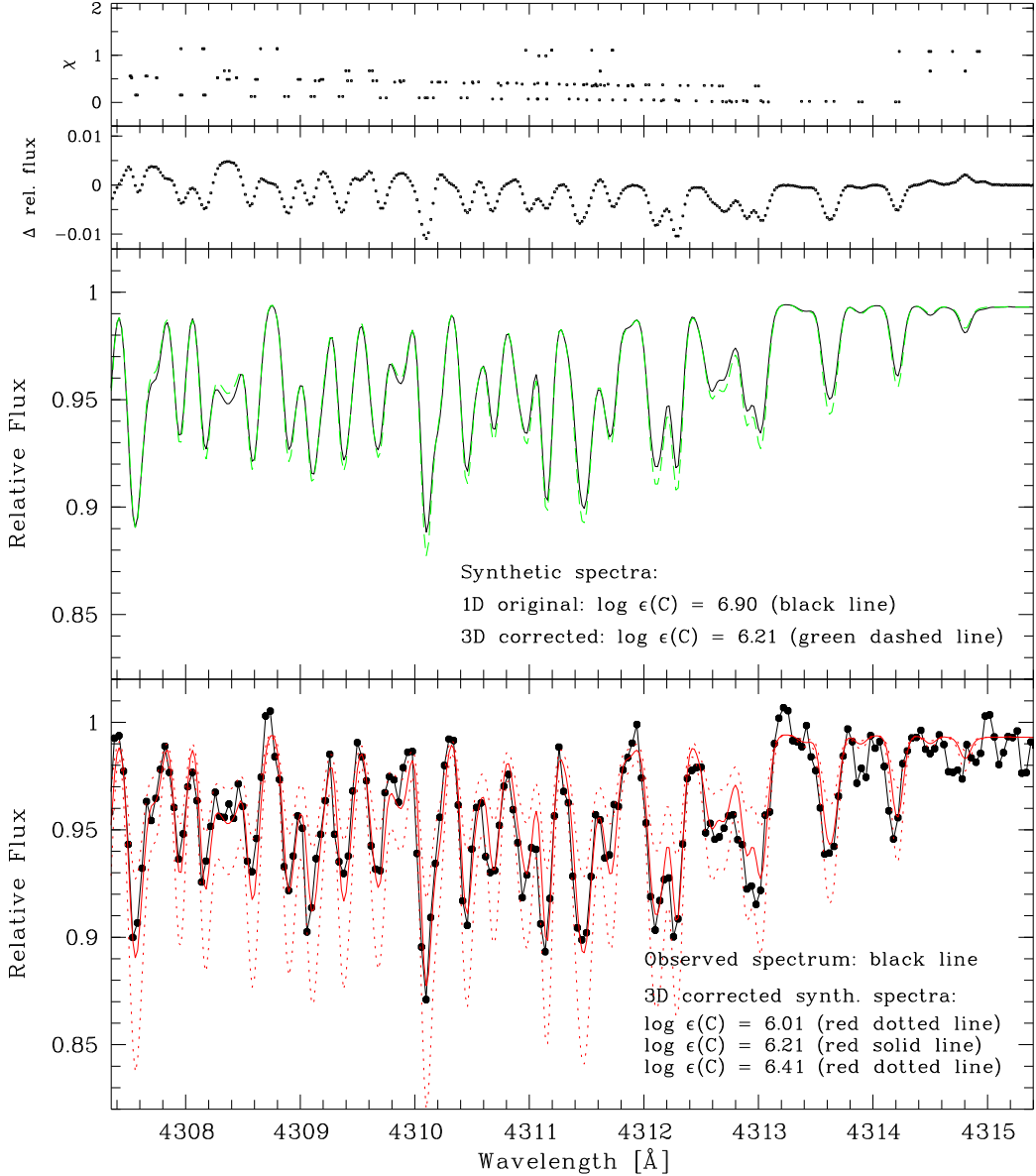}
  \figcaption{\label{CH_comp} CH G-band spectral region of
  HE~1327$-$2326. In the top panel, the excitation potentials of the
  molecular CH lines are presented.  In the second panel from the top,
  the flux difference between the ``1D'' and ``3D-aided'' synthetic spectra
  (from the panel below) is given. A comparison of the ``1D''
  synthetic spectrum with a synthetic spectrum generated from the
  3D-adjusted linelist is shown in the second panel from the bottom. The
  abundance of the ``3D-aided'' synthetic spectrum was varied to reproduce
  the ``1D'' spectrum, and so that the average difference of the two
  fluxes seen in the panel above becomes zero over the given
  wavelength range. Abundances are as given. In the bottom panel we
  show the observed Subaru/HDS spectrum (\textit{dots}) overplotted
  with ``3D-aided'' synthetic spectra with C abundances of
  $\mbox{[C/Fe]}=3.78$ (\textit{solid line}), 3.58, and 3.98
  (\textit{dotted lines}). The average 3D$-$1D correction over this
  wavelength range for CH is $-0.69$\,dex.}
\end{center}
\end{figure}

Before we proceed further, however, a number of tests have to be made:
i) We re-determine the ``1D'' abundances from the new VLT spectrum for
a comparison with previously published values. ii) We synthesize the
spectrum with the new 3D-adjusted linelist.  iii) We compare the
synthetic fit derived from the original ``1D'' linelist with the one
computed with the new 3D-adjusted linelist to explore the influence of
the different 3D$-$1D corrections on different excitation potentials. The
results are shown in Figures~\ref{CH_comp}, \ref{NH_comp}, and
\ref{OH_comp}.

We also caution the reader that the here computed 
average 3D$-$1D abundance corrections are only valid for the
considered wavelength range (because of, e.g., the specific combination of
lower excitation potentials of the contributing lines, the
continuous opacities at these wavelengths, etc.), 
and may thus not apply in general to other molecular features 
in different regions of the spectrum.
Finally, we also remind that any possible 3D non-LTE
or non-equilibrium chemistry effects remain unexplored at this point in time, 
but are expected to be positive \citep{asplund_araa} and therefore
reduce (i.e., make less negative) the effective 3D$-$1D 
correction to CNO abundances.

\subsubsection{Carbon}

The 1D C abundance was already determined from the VLT spectrum by
\citet{o_he1327}. In the meantime, however, a new CH linelist has
become available (B. Plez, priv communication, but see
\citealt{he1523} for some technical details) which yields much
improved syntheses for metal-poor stars. Hence, we use the new
linelist, and derive the C abundance from the G-band head at
$\sim4313$\,{\AA}. The newly derived 1D abundance\footnote{Where $\log
\epsilon ({\rm X}) = \log(N_{\rm X}/N_{\rm H}) + 12.0$} of $\log
\epsilon = 6.90$ agrees very well with previous values of $\log
\epsilon = 6.90$ \citep{HE1327_Nature}, $\log \epsilon = 6.99$
\citep{Aokihe1327} and $\log \epsilon = 6.86$ (from various CH
features; \citealt{o_he1327}).

The excitation potentials of the participating lines in the G-band
spectral region are of a similar level (top panel in the figure), that
results in a almost constant 3D$-$1D correction across that region.
Hence, the variation in abundance between the 1D and 3D-aided
synthetic spectra does not exceed $\sim0.05$\,dex (see the second
panel from the top in Figure~\ref{CH_comp}).  The fit with the
3D-adjusted linelist to the individual lines of the G-band is,
however, slightly improved due to the changes in line strengths
(compare with second panel from the bottom in the figure).  The
overall fit to the data is excellent, and makes it easy to accurately
determine a mean 3D$-$1D correction of C of $-0.69$ (bottom panel of
figure). The final C abundance is $\mbox{[C/Fe]}=3.78$\,dex.  Finally,
concerning the $^{12}$C/$^{13}$C ratio, unfortunately the VLT data do
not cover appropriate wavelengths to re-measure this important ratio.

\subsubsection{Nitrogen} 
The 1D N abundance was determined from the NH band at 3360\,{\AA} for
a comparison with the value obtained from the Subaru discovery
spectrum. Our new fit is much improved due to increased $S/N$ in this
spectral region (now $\sim200$ red-ward of the band). The newly
derived 1D abundance of $\log \epsilon (N)=6.79$ agrees with the
values of 6.68 and 6.83 presented in
\citet{HE1327_Nature}\footnote{For NH, a dissociation energy of 3.47
was used.If a value of 3.37 is employed the abundances becomes $\log
\epsilon (N)=6.78$.} and \citet{Aokihe1327}.

\begin{figure}
 \begin{center}
\includegraphics[height=18cm, clip=true,bbllx=55, bblly=164,
  bburx=526, bbury=714]{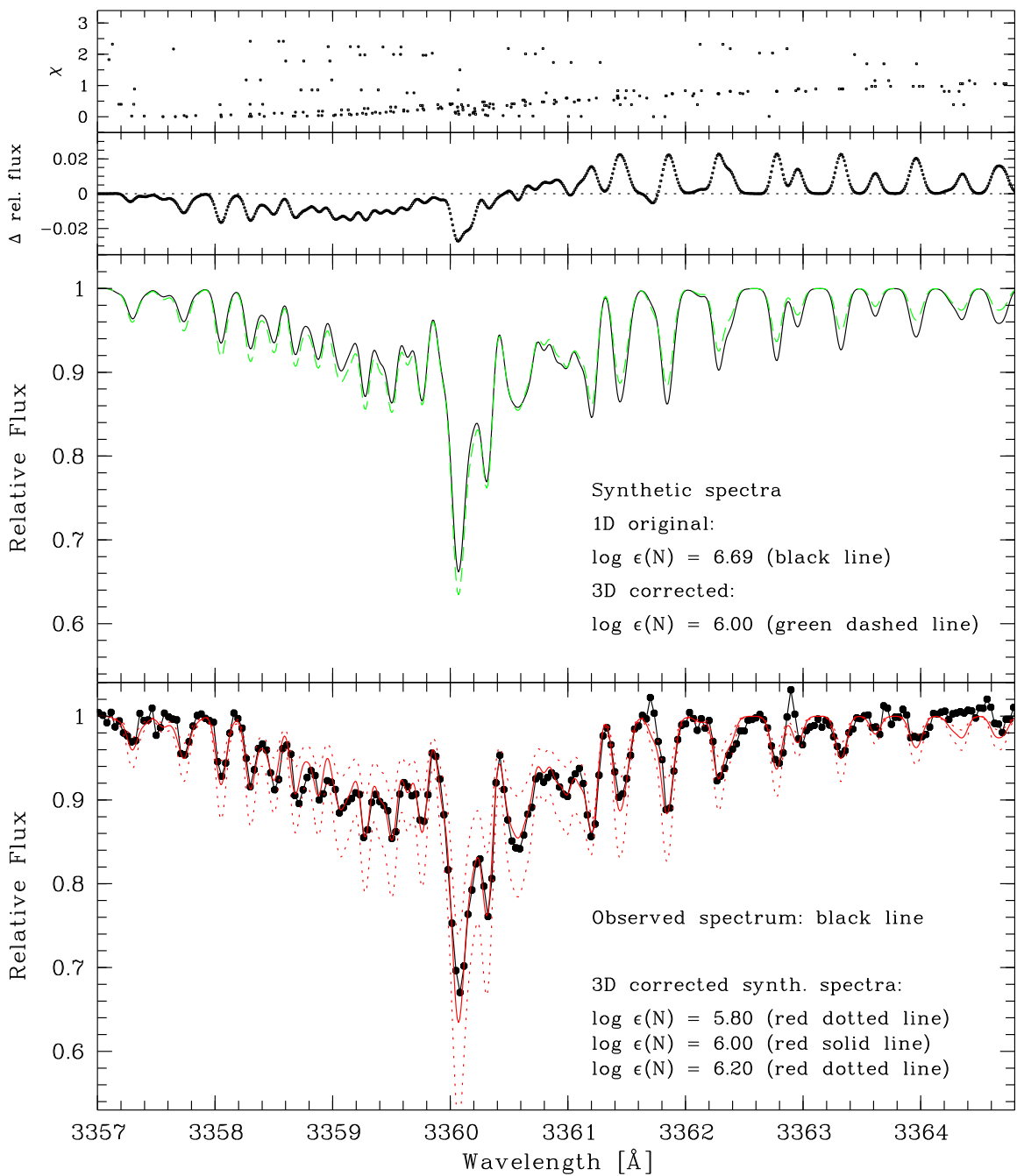}
  \figcaption{\label{NH_comp} Same as for Figure~\ref{CH_comp}, but
  for the NH spectral region. The observed VLT/UVES spectrum (\textit{dots}) in
  the bottom panel is overplotted with ``3D-aided'' synthetic spectra with N
  abundances of $\mbox{[N/Fe]}=4.28$ (\textit{solid line}), 4.08, and
  4.48 (\textit{dotted lines}). The average 3D$-$1D correction over this
  wavelength range for NH is $-0.69$\,dex.}
\end{center}
\end{figure}

\begin{figure}
 \begin{center}
\includegraphics[height=20cm, clip=true,bbllx=52, bblly=84, bburx=526,
  bbury=800]{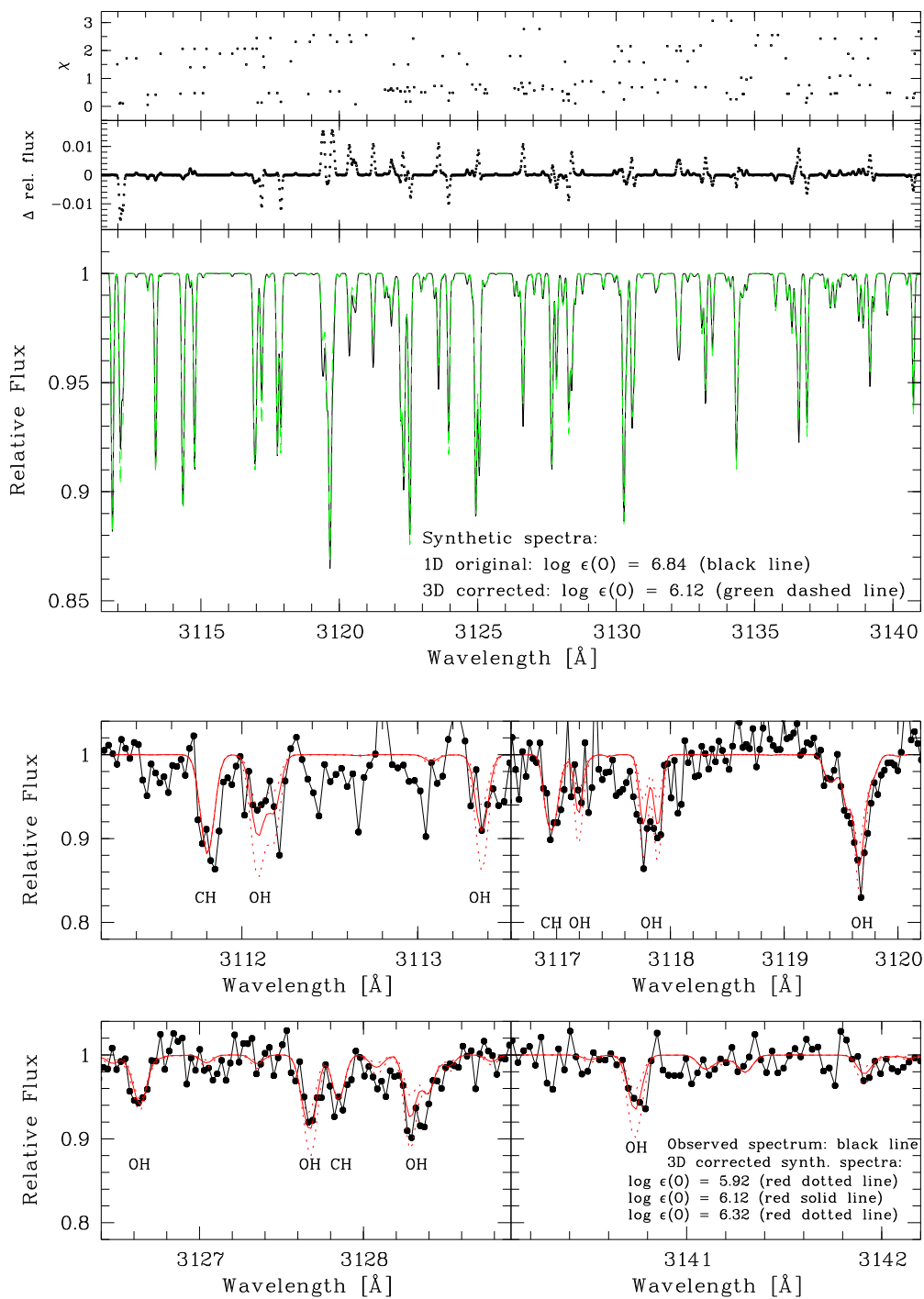}
  \figcaption{\label{OH_comp} Same as for Figure~\ref{CH_comp}, but
  for several spectral region with OH lines. The observed VLT/UVES spectrum
  (\textit{dots}) in the two panels at the bottom is overplotted with
  ``3D-aided'' synthetic spectra with O abundances of $\mbox{[O/Fe]}=3.42$
  (\textit{solid line}), 3.22, and 3.62 (\textit{dotted lines}). The
  average 3D$-$1D correction over this wavelength range for OH is
  $-0.72$\,dex.}
\end{center}
\end{figure}

Figure~\ref{NH_comp} shows the excitation potentials of the molecular
lines that comprise the NH fit. There is a gradual change to higher
excitation potentials from 3357 to 3365\,{\AA}. From the comparison of
the 1D and 3D-aided synthetic spectra it can be seen that the change
in excitation potentials results in the 3D-aided synthetic spectrum
having slightly stronger lines blue-ward ($\sim+0.1$\,dex) of the main
NH feature at 3360\,{\AA} and slightly weaker ($\sim-0.2$\,dex) lines
on the red side (see second panel from top in Figure~\ref{NH_comp}).
Irrespective of any resulting abundance, this effect causes an
improvement of the new fit to the observed NH spectral region compared
with the original 1D linelist fit. 

This effect can not be ``fixed'' with a simple abundance offset since
it would be the same for all participating lines irrespective of
excitation potential. We note that this apparent improvement can not
be accounted to continuum normalization issues since the continuum
points of the synthetic spectra match the observed spectrum very
well. If the continuum of the observed spectrum were determined
incorrectly, those points should not be reproduced by the fit. We
checked the Subaru spectrum, and as can be seen in Figure~7 of Aoki et
al. (2006), the effect of the fit significantly underproducing the
observed spectrum on the blue side, and very slightly overproducing it
on the red side (their best fit is aimed at reproducing the red side)
is present there as well. From this comparison we conclude that it is
not a data reduction effect. We also tested if that effect would
result from the dependence of the 3D$-$1D correction on
line strength, but found it to be negligible. We then investigated the
influence of the effective temperature on relative line strengths,
which in turn depend on their excitation potentials. We changed the
effective temperature of the model atmosphere by $-400$\,K, and fitted
the blue side of the NH band again. On the red side the new fit did
almost match the fit with the original model atmosphere. The
difference on the red side was only about one third of the difference
between the 1D and 3D-aided fits.  It follows that the apparent
improvement of the fit to the data is still present, and can not be
explained with a temperature effect only.

It may thus be possible that the 3D effect can be directly seen in the
NH band, due to a fortuitous distribution of excitation potentials. The
overall 3D$-$1D correction over the NH range is $-0.69$\,dex. A larger, or
smaller, correction would, again, result in a less optimal
reproduction of the observed data, thus suggesting that such a large 3D
correction is correct and necessary to obtain a more accurate N   
abundance for the star. The final N abundance is $\mbox{[N/Fe]}=4.28$.

We note though, that we cannot exclude possible systematic effects in
the present 3D$-$1D differential analysis due to neglected departures from LTE and
from chemical equilibrium at the local temperature. Hypothetically, if
such departures affected all 3D$-$1D corrections for individual molecular
lines (with different excitation potentials) by an approximately
constant factor, then a re-scaling of all gf-values in the 3D-adjusted
linelist by the inverse of the same factor would still lead to a very
good fit of the molecular band but with a different final 3D
correction. The success in the present 3D analysis thus lies in
the fact that the relative line-to-line scatter in the abundances derived from
individual lines in the molecular feature is very much reduced
compared with the 1D analysis.
 
Often, synthetic fits to observed molecular bands are not
fully satisfactory. This is always accounted for in terms of uncertainties in the
molecular line data used to generate the synthetic spectra. While this
may still be true in the case of NH, the uncertainties would be the same for both the
original 1D and the 3D-adjusted linelists, and only affects our conclusions in
the sense that the molecular NH data, i) either is better than
previously thought, or ii) a different, yet unexplored, effect is
responsible for the better fit to the observed data.  The application
of 3D$-$1D corrections in the same way as presented here should be further
investigated for other metal-poor stars to test whether this is a more
general effect or limited to the extreme case of HE~1327$-$2326.  We
note that while the 3D$-$1D corrections for CNO elements HE~0107$-$5240 are
similarly large, the $S/N$ ratio of the data at $\sim3360$\,{\AA} is
not yet sufficient to infer a similar effect for this object (N. Christlieb
et al. 2008, in prep.).

\subsubsection{Oxygen}

The O abundance determination from the VLT data was already described
in detail by \citet{o_he1327}. Apart from the offset in abundance
caused by the application of the 3D$-$1D correction, there generally is
good agreement between the two fits (see the second panel from the top
in Figure~\ref{OH_comp}). The overall 3D$-$1D correction for OH as
determined in this work is $-0.72$\,dex, and thus a slightly lower than what was 
estimated ($-0.9$\,dex) in \citet{o_he1327}. The final O abundance is
$\mbox{[O/Fe]}=3.42$.

\subsection{Lithium and Beryllium}
\begin{figure}
 \begin{center}
\includegraphics[height=8cm, clip=true,bbllx=40, bblly=330, bburx=540,
  bbury=626]{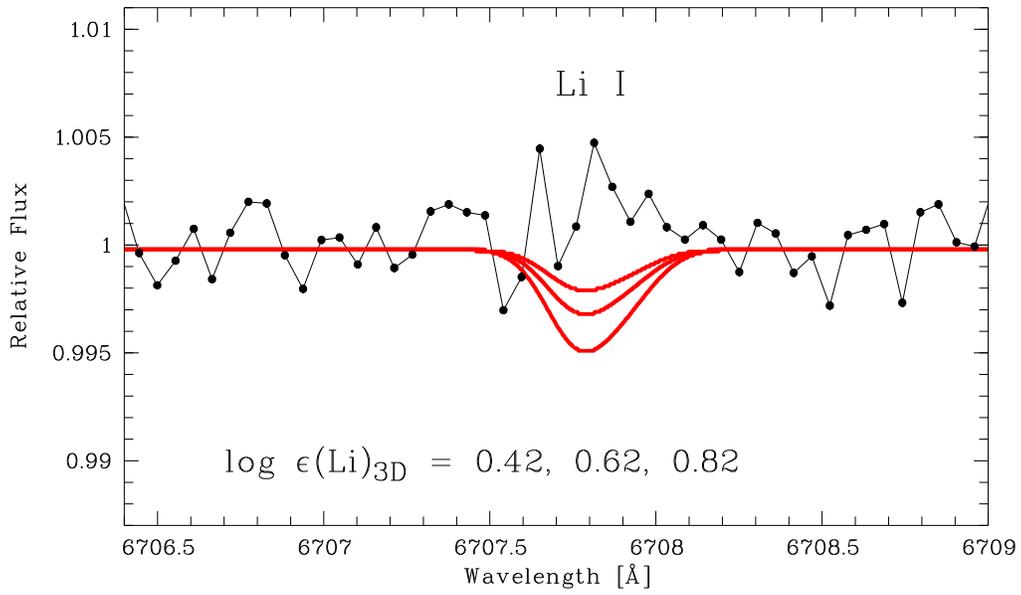} \figcaption{\label{li_plot}
  Spectral region (\textit{connected dots}) of the Li\,I doublet at
  6707\,{\AA} in HE~1327$-$2326. Synthetic spectra are overplotted
  (\textit{solid lines}) for three Li abundances. The upper limit is
  $A({\rm Li})<0.62$.}
\end{center}
\end{figure}

The Li\,I doublet at 6707 {\AA} is {\it still} not detected in this
relatively unevolved star. From the extremely high $S/N$ ($\sim 600$
per pixel) data we are able to derive a new $3\sigma$ upper limit of
$\log \epsilon ({\rm Li}) = A({\rm Li})<0.70$ from our 1D analysis;
this is further reduced to $A({\rm Li})<0.62$ by application of the 3D
correction. Figure~\ref{li_plot} shows the Li spectral regions
overplotted with synthetic spectra of different abundances. This is
significantly lower (i.e., by 1\,dex) than the previous limit of
$\log\epsilon ({\rm Li})=1.6$ \citep{Aokihe1327}. Other unevolved
metal-poor stars with $\mbox{[Fe/H]}\sim -3.5$ have Li values around
$\log\epsilon ({\rm Li})\sim2.0$ (e.g.,
\citealt{ryan_postprim}). These objects have been used to infer the
primordial Li abundance, which can be compared to the value expected
from the baryon-to-photon ratio inferred from WMAP data \citep{WMAP},
$\log\epsilon=2.6$. The discrepancy between the WMAP result and the
observed Li abundance in metal-poor stars has amply been established,
so we shall not further elaborate on it here. We only note that even
though the Li \,I lines, in general, are prone to non-LTE effects,
they are likely to be far too small ($\sim0.1$\,dex) to bridge the gap
between the Li abundance of HE~1327$-$2326 and that of other
metal-poor star as well as the WMAP-inferred value. Also,
gravitational settling of lithium as a possible solution to the
discrepancy is not sufficient \citep{korn_fs3} It is thus very
surprising that the upper limit of HE~1327$-$2326 is not only lower
than the WMAP result but also considerably lower (by more than 1\,dex)
than values derived from other metal-poor star with similar
evolutionary status. This means that either all of this star's Li was
destroyed during its lifetime on the main-sequence, or the star was
born from a Li-depleted material.

\citet{piau} developed a picture where a large fraction of the early
interstellar medium (ISM) is ``recycled'' through massive rotating
Pop\,III stars with the consequence that Li depleted material is
deposited back into the ISM through stellar winds. The first low-mass
stars would primarily form in the wake of those Li-deficient, but CNO
rich winds (\citealt{meynet2005}, but also see above). The new lower
limit of Li (together with high CNO) in HE~1327$-$2326 is in line with
this theory. If this scenario were correct, any (future) observations
of yet-to-be-found dwarfs with metallicities of
$\mbox{[Fe/H]}\lesssim-4.0$ should indeed reveal strong
Li-depletion. Other options for the Li deficiency have been explored
in \citet{Aokihe1327}, and include rapid rotation and HE~1327$-$2326
being a member of a binary system. As for the binary system, continued
radial velocity monitoring does not (yet) suggest the presence of any
companion.

Beryllium is another light, fragile element observed in very
metal-poor turn-off stars. Standard Big-Bang nucleosynthesis models do
not predict a significant yield of this element. Be is destroyed at
$T=3.5 \times 10^{6} $\,K, and is thus slightly more robust than
Li. Hence, if the Be abundance, or a strong upper limit of the
abundance, is determined for HE~1327$-$2326, that could provide
important clues as to why the Li abundance in the star is so extremely
low.

\begin{figure}
 \begin{center}
\includegraphics[height=8cm, clip=true,bbllx=53, bblly=478, bburx=386,
  bbury=677]{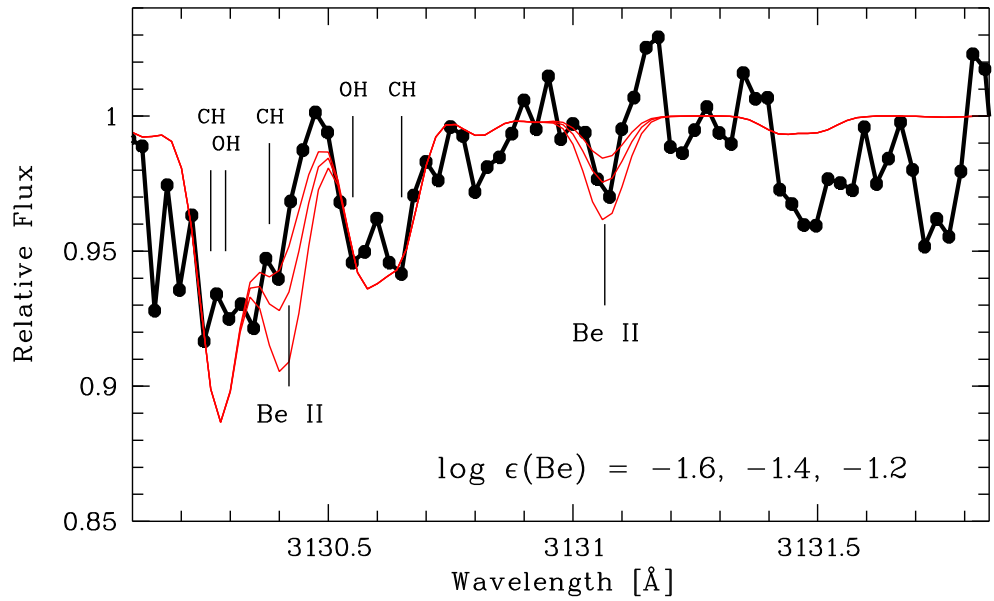} \figcaption{\label{be_plot} Spectral
  region (\textit{connected dots}) of the two beryllium lines at
  $\sim3130$\,{\AA} in HE~1327$-$2326. Synthetic spectra are
  overplotted (\textit{solid lines}) for three Be abundances. The
  upper limit derived from both lines is $\log \epsilon ({\rm Be})<-1.2$.}
\end{center}
\end{figure}

In Figure~\ref{be_plot} we show the spectral region of the two
strongest Be\,II lines in the near UV. As can be seen there is a
(noise) peak at the right wavelength of the $3131$\,{\AA}
line. However, the $S/N$ in this spectral region is low, so that we
cautiously adopt an upper limit of $A({\rm Be})<-1.2$. This value is
as low as the Be abundances measured for extremely metal-poor stars
(e.g., \citealt{boesgaard_novicki}). Thus, we can not derive any
definitive conclusion from the upper limit of the Be abundance for the
depletion of Li in this object.

\subsection{Other elements}
\begin{figure}
 \begin{center}
\includegraphics[height=8cm, clip=true,bbllx=64,
  bblly=342, bburx=270, bbury=540]{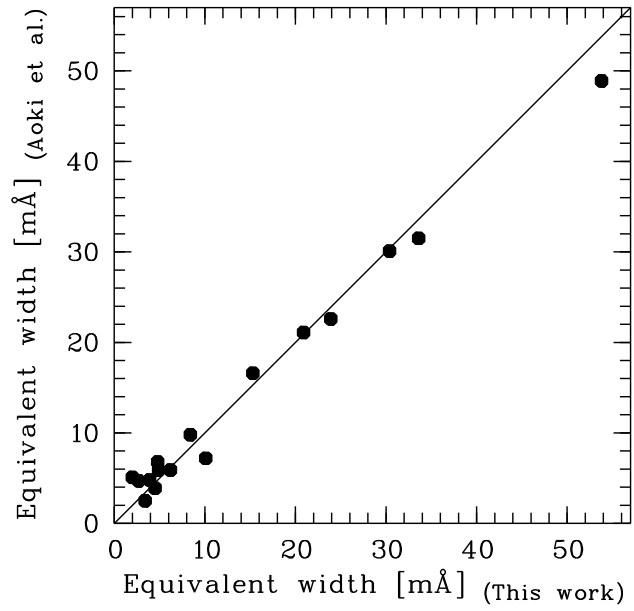}
\figcaption{\label{eqw_comp} Comparison of the equivalent width
measurements between this work and \citet{Aokihe1327}.}
\end{center}
\end{figure}

\begin{figure}
 \begin{center}
\includegraphics[height=4cm, clip=true,bbllx=30,
  bblly=340, bburx=540, bbury=460]{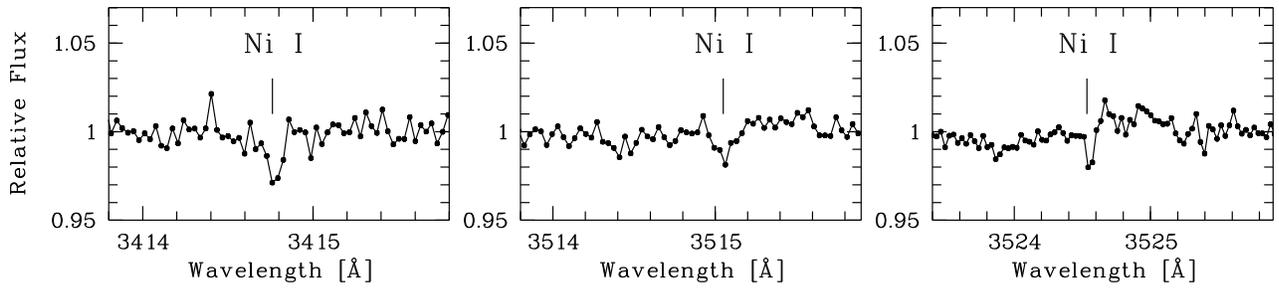}
\figcaption{\label{Ni_plot} Three of the four detected Ni\,I lines in the VLT/UVES spectrum of
 HE~1327$-$2326.}
\end{center}
\end{figure}

For the lines that were previously measured by \citet{Aokihe1327},
there generally is good agreement between their and our equivalent
widths (there is no significant offset). Figure~\ref{eqw_comp} shows a
comparison of the two measurements for the lines in common. We detected
many more lines in the spectrum which confirm the abundances as
derived from the Aoki et al. lines.
We can also report the first detection of four Ni lines in HE~1327$-$2326
(see Figure~\ref{Ni_plot}). Previously only an upper limit could be
derived, but now four weak lines are detected. The equivalent width
measurements and upper limits are presented in Table~\ref{Tab:Eqw},
while the resulting abundances are given in
Table~\ref{Tab:Abundances}.

We also re-determined upper limits for Cr, Co, and Zn. The limits are
much tighter (by $\sim$0.4 to 0.8\,dex) than obtained from the Subaru
spectrum, due to the increased $S/N$ ratio of the VLT data
particularly in the blue spectral region. New upper limits for Be, Sc
and V could also be derived. Our new upper limits, together with the
detection of Ni, are of interest when comparing the overall abundance
pattern with theoretical calculations of SN yields. We will return to
this point in \S~\ref{disc}.

\begin{figure}
 \begin{center}
\includegraphics[height=8cm, clip=true,bbllx=45, bblly=338, bburx=540,
  bbury=626]{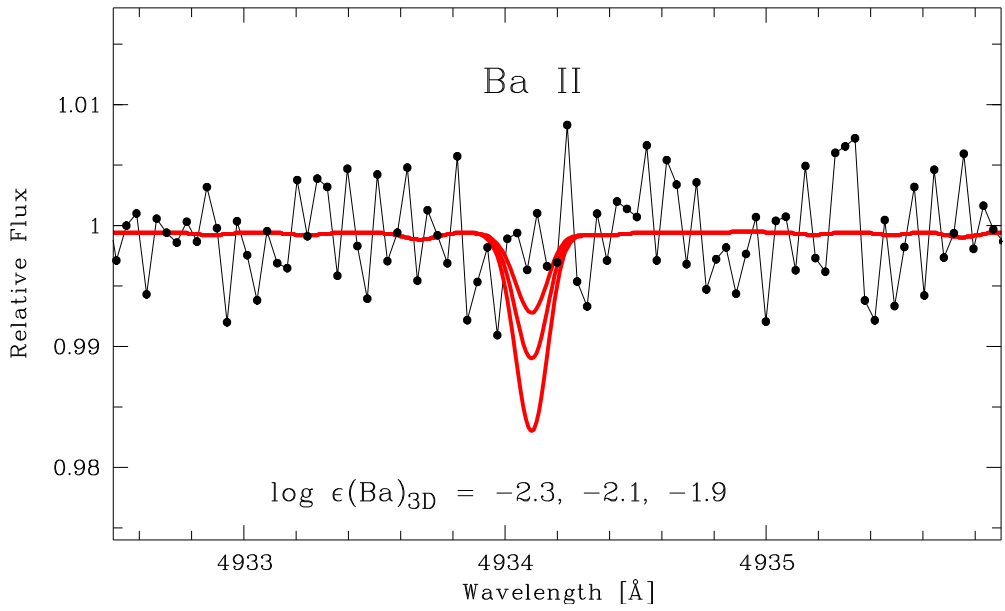} \figcaption{\label{ba_plot}
  Spectral region (\textit{connected dots}) of the barium line at
  4937\,{\AA} in HE~1327$-$2326. Synthetic spectra are overplotted
  (\textit{solid lines}) for three Ba abundances. The upper limit for
  this line is $\log \epsilon ({\rm Ba})<-2.04$.}
\end{center}
\end{figure}

Unfortunately, our spectrum does not cover suitable lines of Mn, Si,
S, Sr, Ba and Eu. That leaves us with no information about Si and S,
and only the previous values from the Subaru spectrum in the case of
Mn, Sr and Ba. For Mn, the second strongest line yields the same upper
limit as has been previously derived from the strongest line at
4020\,{\AA}. The same is the case for Ba. A weaker line at 4943\,{\AA}
line is covered by the red setting. Figure~\ref{ba_plot} shows the Ba
spectral regions overplotted with synthetic spectra of different Ba
abundances. The inferred upper limit yielded almost the same as
was reported in \citet{Aokihe1327} based on the resonance line at
4554\,{\AA}. The resulting ratio is $\mbox{[Sr/Ba]}_{\rm 3D}>-0.29$.

This limit is already higher than what is found in typical r-process
enhanced metal-poor stars ($-0.4$ to $-0.5$;
\citealt{Sneden1998,Hilletal:2002, he1523}). The value is certainly
inconsistent with ratios found in s-process-rich stars, which have
much lower [Sr/Ba] values. Large excesses of light neutron-capture
elements are also found in extremely metal-poor stars, whose abundance
ratios of light to heavy neutron-capture elements, such as Sr/Ba, are
significantly higher than the ratios of the known r-process yields
(e.g., \citealt{aoki05}). The detailed abundance patterns of such
stars have been recently determined by \citet{honda06,honda07}. If the
Sr of HE~1327$-$2326 were related to the process that provided
extremely metal-poor stars with light neutron-capture elements, the Sr
overabundance could be a key to understanding the nucleosynthesis
mechanism of the progenitor object (e.g., \citealt{froehlich}). Such
process is associated with explosive nucleosynthesis, and is thus
compatible with the pre-enrichment scenario by a previous generation
SN as explanation for the overall abundance pattern rather than
through mass transfer across a binary system (see below). As for Eu,
since the strongest line at 4129\,{\AA} is not covered, only a
meaningless upper limit of $\mbox{[Eu/Fe]}<4.1$ could be derived from
a line at 6645\,{\AA}.

For completeness, we also provide the upper limits derived from the
O\,I triplet and forbidden O\,I lines, as presented in
\citet{o_he1327}. Upper limits on the equivalent width and abundances
are listed in Tables~\ref{Tab:Eqw}~and~\ref{Tab:Abundances}.

\section{Discussion and Conclusions}\label{disc}
We have presented an abundance analysis of the currently most Fe-poor
star, HE~1327$-$2326, based on the currently best available
observational data obtained with VLT/UVES. In an attempt to obtain the
best possible abundances from those data, we have combined our 1D
abundances with state-of-the-art 3D$-$1D model atmosphere
corrections. Hence, these abundances supersede our previous
measurements.  We recommend using these new 3D$-$1D corrected LTE
abundances for future modeling of the abundance pattern of
HE~1327$-$2326.

\begin{figure}
 \begin{center}
\includegraphics[height=12cm, clip=true,bbllx=45, bblly=88, bburx=540,
  bbury=426]{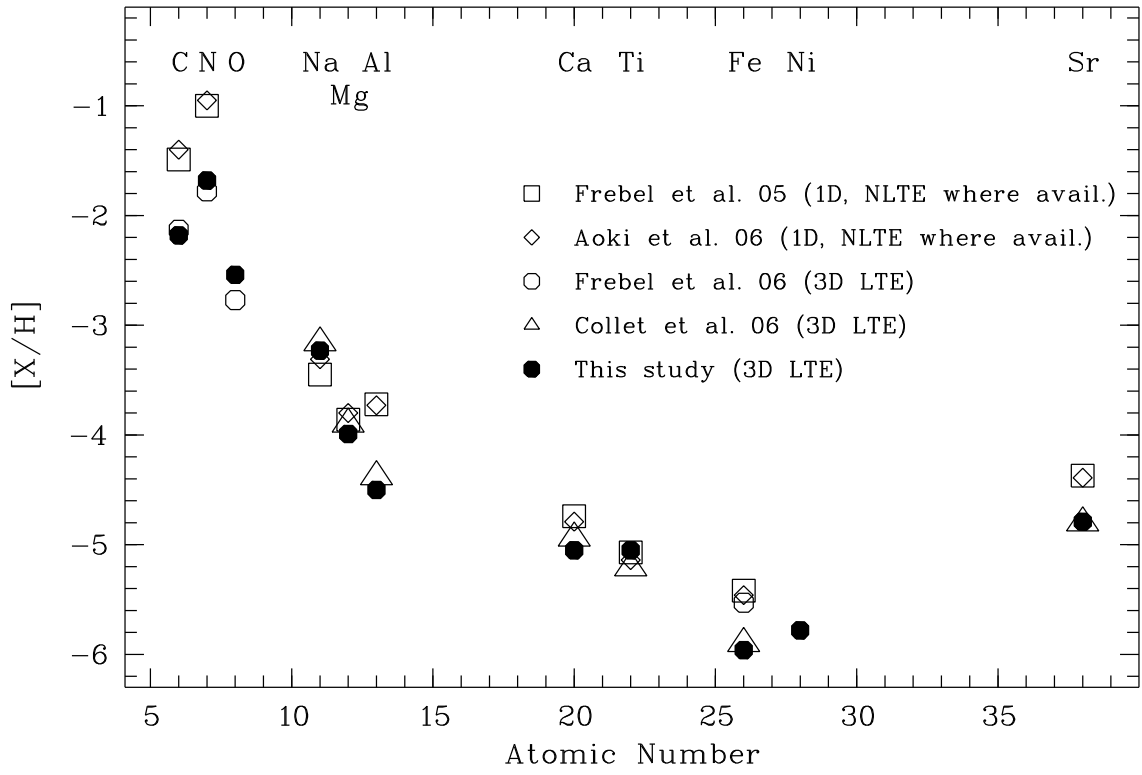} \figcaption{\label{abund_summ} Comparison of
  previously published abundances [X/H] of HE~1327$-$2326.}
 \end{center}
\end{figure}

Figure~\ref{abund_summ} summarizes all abundances measurements for the
star published to date. We plot the ``best available'' values at the
time, i.e., 1D abundances with non-LTE corrections where available in
the older cases, and LTE abundances with differently determined 3D
corrections in the newer studies.  The Fe abundance differs in these
works, so that we compare [X/H] values which leave Fe as ``free
parameter''. The 3D$-$1D corrected abundances are somewhat shifted
towards lower values compared with the 1D abundances. The biggest
differences are found among the CNO elements ($\Delta\sim-0.7$)
whereas for other elements the differences between 1D and 3D analyses
are less pronounced. We note that when comparing [X/Fe] measurements,
because of the different [Fe/H] values, some of those differences in
[X/H] are canceled out. This fact should be kept in mind when
comparing the overall abundance pattern with theoretically derived
abundances predictions based on SN yields.

\begin{figure}
 \begin{center}
\includegraphics[height=12cm, clip=true,bbllx=45, bblly=408, bburx=540, bbury=726]{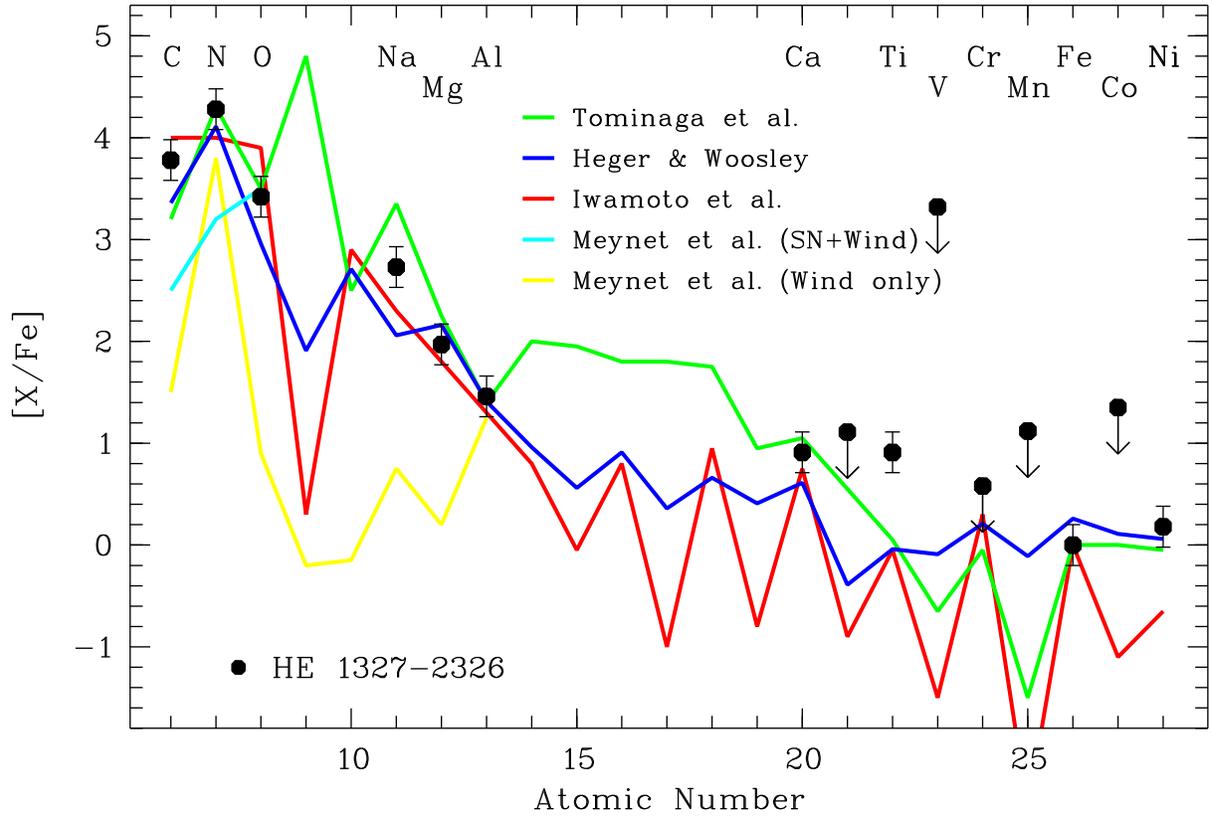} 
  \figcaption{\label{models}
  Comparison of the new 3D$-$1D corrected abundances [X/Fe] of
  HE~1327$-$2326 with predictions from various models based on
  chemical yields of the first-generation SNe. See text for
  discussion.}
 \end{center}
\end{figure}

Several groups of authors have already employed the (in the most cases
the old) abundances to test current scenarios for the chemical yields
of the first-generation SNe. In order to compare how well all these
predictions reproduce our new 3D$-$1D corrected abundances presented
here, we plot them all in Figure~\ref{models}.  We note that all the
above interpretations assume that the star is not a member of a binary
system. This is indeed supported by ongoing radial velocity
measurements of HE~1327$-$2326. Two measurements, one from 2007
(D. Lai, priv. comm.) and one from 2008 agree with the previous data
\citep{o_he1327} and does not indicate any significant change of the
radial velocity.

\citet{iwamoto_science} were the first authors to compare the 1D
abundances of HE~1327$-$2326 \citep{HE1327_Nature} with their yields
of a 25\,M$_{\odot}$ Pop\,III hypernova, for given mixing and fallback
parameters. Particularly, the high CN abundance together with the
elevated Mg abundance were well reproduced by their yields. An updated
fit was presented in \citet{tominaga07} after the O abundance (with
estimates for the 3D$-$1D correction) of HE~1327$-$2326 had been determined
\citep{o_he1327}. Hence, this model now particularly well fits our new CNO
abundances. Na, Mg, Al, Ca and Fe abundances are well reproduced,
too. The new abundance of Ni (3D$-$1D corrected) also agrees with
their predicted Ni yield.

\citet{meynet2005} investigated mass loss by a massive
(60\,M$_{\odot}$) rotating Pop\,III star, and the chemical yields of
the subsequent SN explosion. They find that from the combined yields
of the stellar wind and the SN, the highly elevated 1D CNO pattern of
HE~1327$-$2326 \citep{HE1327_Nature} and that of other metal-poor
stars can qualitatively be reproduced. In their ``wind-only'' model, a
high N abundance can be achieved, although all other elements
considered (except Al) are produced in insufficient quantities to
reproduce the observed abundance pattern.

\citet{heger2008} have computed new chemical yields for a range
(10--100\,M$_{\odot}$) of Pop\,III SNe. They are fitting the observed
abundances of HE~1327$-$2326 \citep{Aokihe1327} with either the yield
of a single SN or the integrated yields of a range of SNe for a given
initial mass function. For various progenitor masses (mass ranges),
explosion energies, and mixing parameters the abundance pattern can be
roughly reproduced. The best fit model for a single SN requires a
21.5\,M$_{\odot}$ progenitor, while the best fit model for integrated
yields stems from the range 15-35\,M$_{\odot}$ of
progenitors. Overall, this is a similar progenitor mass as
\citet{iwamoto_science} suggested. In Figure~\ref{models} we show the
``best single stars'' fit (Panel A in their Figure~14;
\citealt{heger2008}) in comparison with our new abundances. Overall,
the fit is in reasonable agreement with the data and provides an
interesting counterpart to the work of \citet{tominaga07}.

Recently, \citet{dtrans} used C and O abundances of metal-poor stars
for a comparison with theoretical predictions for the critical
metallicity that is required to facilitate low-mass star formation in
the early Universe. Based on the theory of fine-structure line
cooling, it was predicted that all stars with
$\mbox{[Fe/H]}\lesssim-4.0$ should have elevated C and/or O abundances
and have $D_{\rm trans}=\log(10^{\rm{[C/H]}}+10^{\rm{[O/H]}}) >-3.5$\,dex. The
new 3D$-$1D corrected abundances of HE~1327$-$2326 presented here lead
to $D_{\rm trans}=-2.18$\,dex, well above the critical value.

\section{Outlook}\label{outlook}

To improve our understanding of the conditions of the early Universe,
and to put all the above theories to further critical observational
tests, clearly, more stars are needed with metallicities of
$\mbox{[Fe/H]}\lesssim-4$. \citet{he0557} recently reported the
discovery of a giant with $\mbox{[Fe/H]}=-4.75$. In particular, more
giants should be sought at the lowest metallicities because their
(lower) temperature, at a given abundance, allows for stronger lines
that can be detected more easily.  The Fe lines in the subgiant
HE~1327$-$2326 are only barely detectable because of its hotter
temperature. Further ultra-metal-poor stars to be discovered in the
future will likely be fainter than HE~1327$-$2326 ($B\sim14.0$\,mag)
and even HE~0107$-$5240 ($B\sim15.6$\,mag), which imposes yet another
challenge when it comes to obtaining the necessary high $S/N$ spectra to
measure very weak spectral lines particularly in the blue spectral
region. Detailed studies of larger samples of stars at
$\mbox{[Fe/H]}<-5.0$ will therefore require 30\,m class telescopes,
which will make possible such difficult but extremely important
observations.

\acknowledgements 
We thank Martin Asplund for discussions on abundance
analyses based on 3D model stellar atmospheres and an anonymous
referee for valuable suggestions regarding our new 3D$-$1D abundance
correction technique.  A.~F. thanks Uppsala Astronomical Observatory
for its hospitality during parts of the write-up of this paper. She
acknowledges support through the W.~J.~McDonald Fellowship of the
McDonald Observatory. N.~C. acknowledges financial support by Deutsche
Forschungsgemeinschaft through grants Ch~214/3 and Re~353/44. He is a
Research Fellow of the Royal Swedish Academy of Sciences supported by
a grant from the Knut and Alice Wallenberg Foundation.

\textit{Facilities:} 
VLT:Kueyen (UVES)

\clearpage
\begin{deluxetable}{lclcc} 
\tablecolumns{9} 
\tablewidth{0pt} 
\tabletypesize{\small}
\tablecaption{\label{Tab:SN} Measured $S/N$ ratios in different wavelength regions}
\tablehead{
\colhead{$\lambda$ ({\AA}) } & 
\colhead{$S/N$ (pixel)}    & 
\colhead{Setting}  & 
\colhead{Pixel size (km\,s$^{-1}$)}    &
\colhead{FWHM of Th-Ar lines (pixel)}    }
\startdata

$\sim 3500$ &$\sim 170$ & BLUE 346\,nm& $\sim 2.7$ & $\sim 3.8$  \\
$\sim 3800$ &$\sim 250$ & BLUE 346\,nm& $\sim 2.4$ & $\sim 3.7$  \\
$\sim 5000$ &$\sim 470$ & RED 580\,nm & $\sim 2.3$ & $\sim 3.2$  \\
$\sim 6000$ &$\sim 750$ & RED 580\,nm & $\sim 2.4$ & $\sim 3.5$  \\
$\sim 6700$ &$\sim 600$ & RED 580\,nm & $\sim 2.1$ & $\sim 3.5$  \\
$\sim 8500$ &$\sim 200$ & RED 760\,nm & $\sim 2.3$ & $\sim 3.4$    
\enddata
\end{deluxetable} 

\clearpage
\begin{deluxetable}{lccrrrrrl}
\tabletypesize{\footnotesize}
\tabletypesize{\tiny}
\tablewidth{0pc} 
\tablecaption{\label{Tab:Abundances} VLT/UVES abundances of HE~1327$-$2326}
\tablehead{ 
\colhead{Species} & 
\colhead{$\log\epsilon (\mbox{X})_{\odot}$} & \colhead{$N_{\mbox{\scriptsize lin.}}$\tablenotemark{a}} &
\colhead{$\log\epsilon (\mbox{X})_{\rm 1D}$} &
\colhead{$\log\epsilon (\mbox{X})_{\rm 3D}$} &
\colhead{[X/H]$_{\rm 3D}$} & \colhead{[X/Fe]$_{\rm 3D}$} }
\startdata          
C\,(CH)& $8.39$ &syn    & $6.90 $ & $6.21$   & $-2.18$ & $3.78$ \\
N\,(NH)& $7.78$ &syn    & $6.79 $ & $6.10$   & $-1.68$ & $4.28$ \\
O\,(OH)& $8.66$ &syn    & $6.84 $ & $6.12$   & $-2.54$ & $3.42$ \\
Na\,I  & $6.17$ &2      & $2.99 $ & $2.94$   & $-3.23$ & $2.73$ \\ 
Mg\,I  & $7.53$ &4      & $3.54 $ &  3.54    & $-3.99$ & $1.97$ \\ 
Al\,I  & $6.37$ &1      & $1.90 $ &  1.87    & $-4.50$ & $1.46$ \\ 
Ca\,I  & $6.31$ &1      & $0.88 $ &  0.79    & $-5.52$ & $0.44$ \\ 
Ca\,II & $6.31$ &4      & $1.34 $ &  1.26    & $-5.05$ & $0.91$ \\ 
Ti\,II & $4.90$ &15     & $-0.09$ &$-0.15$   & $-5.05$ & $0.91$ \\ 
Fe\,I  & $7.45$ &12     & $ 1.79$ &  1.49    & $-5.96$ &\nodata \\ 
Ni\,I  & $6.23$ &4      & $ 0.73$ & $ 0.45$  & $-5.78$ & $0.18$ \\ 
Sr\,II & $2.92$ &2      & $-1.76$ & $-1.87$  & $-4.79$ & $1.17$ \\ 
\hline				    				     	
Li\,I  & $1.05$ & 6707  & $<0.70$ & $ <0.62$ & \nodata  &\nodata  \\ 
Be\,II & $1.38$ & 3131  & $<-1.20$& $<-1.15$ & \nodata  &\nodata  \\
Sc\,II & $3.05$ & 3613  & $<-1.68$& $<-1.80$ & $<-4.85$ & $<1.11$ \\ 
V\,I   & $4.00$ & 3184  & $<1.50$ & $ <1.36$ & $<-2.64$ & $<3.32$ \\ 
Cr\,I  & $5.64$ & 3579  & $<0.45$ & $ <0.26$ & $<-5.38$ & $<0.58$ \\ 
Mn\,I  & $5.39$ & 3439  & $<0.84$ & $ <0.55$ & $<-4.84$ & $<1.12$ \\ 
Fe\,II & $7.45$ & 3227  & $<1.99$ & $ <2.05$ & $<-5.40$ & $<0.56$ \\ 
Co\,I  & $4.92$ & 3454  & $<0.58$ & $ <0.31$ & $<-4.61$ & $<1.35$ \\ 
Zn\,I  & $4.60$ & 4810  & $<1.61$ & $ <1.65$ & $<-2.95$ & $<3.01$ \\ 
Ba\,II & $2.17$ & 4554  & $<-2.14$& $<-2.39$ & $<-4.56$ & $<1.40$ \\ 
Eu\,II & $0.52$ & 6645  & $<-0.76$& $<-0.80$ & $<-1.32$ & $<4.64$ \\ 
\enddata 
\tablecomments{
All [X/Fe] 3D ratios are computed with $\mbox{[Fe/H]}=-5.96$. Solar abundances
have been taken from \citet{solar_abund}.}

\tablenotetext{b}{``Syn'' indicates the use of spectrum synthesis for
the abundance determination. For upper limits the wavelength of the
employed line is given.}
\tablenotetext{b}{3D CNO corrections have been determined from
spectrum synthesis, see text for details.}  

\end{deluxetable}

\clearpage
\begin{deluxetable}{lllrrrrr} 
\tablecolumns{9} 
\tablewidth{0pt} 
\tabletypesize{\small}
\tablecaption{\label{Tab:Eqw} Atomic data and measured equivalent widths}
\tablehead{
  \colhead{} & \colhead{} & \colhead{} & \colhead{} & \multicolumn{2}{c}{$W_{\lambda}$}\\
  \cline{5-6}\\
  \colhead{}    & \colhead{$\lambda$} & \colhead{$\chi$} & \colhead{$\log gf$} &
  \colhead{This work} & \colhead{Aoki et al.} \\
  \colhead{Species} & \colhead{({\AA})}    & \colhead{(eV)}  & \colhead{(dex)}  &
  \colhead{(m{\AA})} & \colhead{(m{\AA})}  &\colhead{$\log \epsilon$(1D)} &\colhead{$\log \epsilon$(3D)}
  }
\startdata
Na\,I & $5889.951$ & $0.00$& $ 0.12$ &  53.8 & $ 48.9$  & 2.99   &$2.93$ \\
Na\,I & $5895.924$ & $0.00$& $-0.18$ &  33.6 & $ 31.5$  & 2.99   &$2.94$ \\
Mg\,I & $3829.355$ & $2.71$& $-0.21$ &  23.9 & $ 22.6$  & 3.48   &$3.50$ \\
Mg\,I & $5167.321\tablenotemark{a}$ & $2.71$& $-1.03$ &   8.4 & $  9.8$  & 3.68   &$3.68$ \\ 
Mg\,I & $5172.684$ & $2.71$& $-0.40$ &  20.9 & $ 21.1$  & 3.50   &$3.49$ \\
Mg\,I & $5183.604$ & $2.72$& $-0.18$ &  30.4 & $ 30.1$  & 3.50   &$3.49$ \\
Al\,I & $3961.529$ & $0.01$& $-0.34$ &      b& $ 11.0$  & 1.92   &$1.87$ \\
Ca\,I & $4226.728$ & $0.00$& $ 0.24$ &      b& $  2.7$  & 0.89   &$0.79$ \\
Ca\,II& $3933.663$ & $0.00$& $ 0.11$ &      b& $128.9$  & 1.31   &$1.16$ \\
Ca\,II& $8498.023$ & $1.69$& $-1.42$ &$ 10.0$& \nodata  & 1.31   &$1.28$ \\
Ca\,II& $8542.091$ & $1.70$& $-0.46$ &$ 54.7$& \nodata  & 1.38   &$1.29$ \\
Ca\,II& $8662.141$ & $1.69$& $-0.72$ &$ 39.1$& \nodata  & 1.36   &$1.31$ \\
Ti\,II&  3072.984  &  0.00 & $-0.36$ &  5.4  &\nodata   &$ 0.20$ &$ 0.11$ \\
Ti\,II&  3088.037  &  0.05 & $ 0.25$ &  7.2  &\nodata   &$-0.24$ &$-0.32$ \\
Ti\,II&  3229.198  &  0.00 & $-0.55$ &  3.5  &\nodata   &$ 0.17$ &$ 0.08$ \\
Ti\,II& $3234.520$ & $0.05$& $ 0.43$ & 10.1  & $  7.2$  &$-0.27$ &$-0.34$ \\   
Ti\,II&  3236.578  &  0.03 & $ 0.23$ &  6.8  &\nodata   &$-0.29$ &$-0.38$ \\
Ti\,II&  3239.044  &  0.01 & $ 0.06$ &  7.5  &\nodata   &$-0.08$ &$-0.17$ \\
Ti\,II&  3241.994  &  0.00 & $-0.05$ &  3.9  &\nodata   &$-0.30$ &$-0.39$ \\
Ti\,II&  3322.941  &  0.15 & $-0.09$ &  3.8  &\nodata   &$-0.13$ &$-0.22$ \\
Ti\,II& $3349.408$ & $0.05$& $ 0.59$ & 15.3  & $ 16.6$  &$-0.23$ &$-0.33$ \\
Ti\,II&  3372.800  &  0.01 & $ 0.27$ & 13.5  &\nodata   &$-0.02$ &$-0.12$ \\
Ti\,II&  3380.279  &  0.05 & $-0.57$ &  4.8  &\nodata   &$ 0.36$ &$ 0.27$ \\
Ti\,II&  3383.768  &  0.00 & $ 0.14$ &  9.1  &\nodata   &$-0.10$ &$-0.20$ \\
Ti\,II&  3387.846  &  0.03 & $-0.43$ &  5.1  &\nodata   &$ 0.23$ &$ 0.13$ \\
Ti\,II& $3759.296$ & $0.61$& $ 0.27$ & 4.9   & $  5.9$  &$-0.04$ &$-0.08$ \\
Ti\,II& $3761.323$ & $0.57$& $ 0.17$ & 2.7   & $  4.7$  &$-0.24$ &$-0.28$ \\
Fe\,I & $3440.606$ & $0.00$& $ -0.67$&$ 5.1$ &\nodata   & 1.86   &$1.48$ \\
Fe\,I &  3440.989  &  0.05 & $ -0.96$&$ 3.0$ &\nodata   & 1.80   &$1.43$ \\
Fe\,I & $3581.193$ & $0.86$& $ 0.42$ &$ 6.2$ & $  5.9$  & 1.75   &$1.56$ \\
Fe\,I & $3618.768$ & $0.99$& $ 0.00$ &  1.2: &\nodata   & 1.53:\tablenotemark{c}&$1.36$ \\
Fe\,I & $3719.935$ & $0.00$& $ -0.43$&$ 7.9$ &\nodata   & 1.85   &$1.48$ \\
Fe\,I & $3737.131$ & $0.05$& $-0.57$ &$ 4.5$ & $  3.9$  & 1.76   &$1.41$ \\
Fe\,I & $3745.561$ & $0.09$& $-0.77$ &$ 3.9$ & $  4.8$  & 1.82   &$1.48$ \\
Fe\,I & $3748.262$ & $0.11$& $ -1.02$&$ 1.9$ &\nodata   & 1.86   &$1.53$ \\
Fe\,I & $3758.233$ & $0.96$& $-0.01$ &$ 2.0$ & $  5.1$  & 1.69   &$1.54$ \\
Fe\,I & $3820.425$ & $0.86$& $ 0.16$ &$ 3.4$ & $  2.5$  & 1.66   &$1.49$ \\
Fe\,I & $3825.881$ &  0.92 & $-0.04$ &  1.8: &\nodata   & 1.58:\tablenotemark{c}&$1.43$ \\
Fe\,I &  3856.372  &  0.05 & $-1.29$ &  1.1: &\nodata   & 1.79:\tablenotemark{c}&$1.43$ \\
Fe\,I & $3859.912$ & $0.00$& $-0.71$ &  4.8  & $  6.8$  & 1.83   &$ 1.45$ \\
Fe\,I & $4045.812$ & $1.49$& $ 0.29$ &     b & $  1.9$  & 1.78   &$ 1.68$ \\ 
Ni\,I & $3414.761$ & $0.03$& $-0.03$ & $ 2.9$& $<7.5$   &$ 0.70$  &$ 0.40$ \\ 
Ni\,I &  3515.049  &  0.11 &$-0.21 $ &   2.1 & \nodata  &  0.81   &$ 0.53$ \\
Ni\,I & $3524.535$ & $0.03$& $ 0.01$ &  $2.4$& \nodata  &  0.57   &$ 0.26$ \\ 
Ni\,I &  3619.386  &  0.42 &   0.04  &   2.0 & \nodata  &  0.83   &$ 0.61$ \\  
Sr\,II& $4077.724$ & $0.00$& $ 0.16$ &      b& $  7.3$  &$-1.76$  &$-1.87$ \\
Sr\,II& $4215.540$ & $0.00$& $-0.16$ &      b&    3.8   &$-1.75$  &$-1.86$ \\
\hline
Li\,I &  6707.761  &  0.00 & $-0.01$& syn\tablenotemark{d}&\nodata&$<0.70$ &$0.62$ \\
Li\,I &  6707.912  &  0.00 & $-0.31$& syn\tablenotemark{d}&\nodata&$<0.70$ &$0.62$ \\
Be\,II&  3130.420  &  0.00 & $-0.17$& syn\tablenotemark{d}&\nodata&$<-1.20$&$-1.15$ \\
Be\,II&  3131.065  &  0.00 & $-0.47$& syn\tablenotemark{d}&\nodata&$<-1.20$&$-1.15$ \\
\mbox{[O\,I]}&  6300.304  &  0.00 & $-9.82$& $<1.8$&\nodata & $<8.01$   &$< 8.07$ \\
O\,I  &  7771.944  &  9.15 & $ 0.32$& $<2.0$&\nodata & $<6.19$   &$< 6.21$ \\
O\,I  &  7774.166  &  9.15 & $ 0.17$& $<2.0$&\nodata & $<6.34$   &$< 6.36$ \\
O\,I  &  7775.388  &  9.15 & $-0.05$& $<2.0$&\nodata & $<6.56$   &$< 6.58$ \\
Sc\,II& $3572.526$ & $0.02$& $ 0.27$ &$<1.1$& \nodata& $<-1.57$  &$<-1.68$ \\
Sc\,II& $3613.829$ &  0.02 &   0.42  &$<1.0$& \nodata& $<-1.68$  &$<-1.80$ \\
Sc\,II& $3642.784$ & $0.00$& $ 0.13$ &$<1.2$& \nodata& $<-1.45$  &$<-1.54$ \\
V\,I  &  3183.410  &  0.02 &   0.46  &$<2.6$& \nodata& $<1.60$   &$< 1.46$ \\ 
V\,I  &  3183.970  &  0.04 &   0.58  &$<2.6$& \nodata& $<1.50$   &$< 1.36$ \\ 
V\,I  &  3184.013  &  0.00 &   0.34  &$<2.6$& \nodata& $<1.70$   &$< 1.54$ \\
Cr\,I & $3578.684$ & $0.00$& $ 0.41$ &$<1.2$& \nodata& $<0.45$   &$< 0.26$ \\
Cr\,I & $3593.481$ & $0.00$& $ 0.31$ &$<1.3$& \nodata& $<0.59$   &$< 0.40$ \\
Cr\,I & $4254.332$ & $0.00$& $-0.11$ &     b& $<2.0$ & $<1.07$   &$< 0.88$ \\ 
Mn\,I & $4030.753$ & $0.00$& $-0.47$ &     b& $<2.0$ & $<0.84$   &$< 0.55$ \\
Fe\,II& $3227.742$ & $1.67$& $-1.10$ &$<2.2$& \nodata& $<1.99$   &$< 2.05$ \\ 
Fe\,II& $5018.450$ & $2.89$& $-1.22$ &$<0.8$& $<2.0$ & $<2.61$   &$< 2.66$ \\ 
Co\,I & $3453.514$ & $0.43$& $ 0.38$ &$<1.5$& $<6.3$ & $<0.58$   &$< 0.31$ \\
Zn\,I & $4810.528$ & $4.08$& $-0.15$ &$<1.0$& $<2.0$ & $<1.61$   &$< 1.65$ \\
Ba\,II& $4554.029$ & $0.00$& $ 0.17$ &    b & $<1.8$ & $<-2.14$  &$<-2.39$ \\
Ba\,II& $4934.100$ & $0.00$& $-0.16$ &$<1.1$& \nodata& $<-2.04$  &$<-2.23$ \\
Eu\,II& $6645.064$ & $1.38$& $ 0.21$ &$<1.1$& \nodata& $<-0.76$  &$<-0.80$
\enddata
\tablenotetext{a}{The blend with an Fe line is negligible.}
\tablenotetext{b}{These lines are not covered by the new VLT data.}
\tablenotetext{c}{These lines are not included in the final abundance estimate.}
\tablenotetext{d}{Abundance limit was derived from spectrum synthesis.}
\end{deluxetable} 

\clearpage

\clearpage

\clearpage

\clearpage

\clearpage

\clearpage

\clearpage

\clearpage

\clearpage

\clearpage

\clearpage 

\clearpage                                     

\clearpage 

\clearpage                                                                

\clearpage   

\clearpage

\clearpage

\end{document}